\documentclass[12pt]{article}
\usepackage[a4paper,margin=0.75in]{geometry}
\usepackage[onehalfspacing]{setspace}

\usepackage{mathptmx}
\usepackage{amsmath,amssymb,amsthm}

\usepackage{pdflscape}
\usepackage{graphicx}
\usepackage{epstopdf}
\usepackage{caption}
\usepackage{subcaption}

\usepackage{natbib}
\usepackage{chngcntr}
\usepackage{hyperref}

\begin{document}

\title{Income Taxes, Gross Hourly Wages, and the Anatomy of Behavioral Responses: Evidence from a Danish Tax Reform\thanks{We would like to thank Mike Brewer, Pierre Cahuc, Manasi Deshpande, Peter Fredriksson, Chishio Furukawa, Nikolaj A. Harmon, Bo Honor{\'e}, Niels Johannesen, Daiji Kawaguchi, Francis Kramarz, Claus T. Kreiner, Jakob R. Munch, Fabien Postel-Vinay, Andrew Shephard, Oskar N. Skans, Linh T. T\^{o}, Shintaro Yamaguchi, and seminar/conference participants at Aarhus University, CREST, EALE, IIPF, Osaka University, RIETI, Royal Holloway, and ZEW for their comments. Financial support from Aarhus University, RIETI, and Royal Holloway is gratefully acknowledged. Sumiya would also like to thank AUFF and, in particular, the Associate Professor Starting Grant awarded to Rune Vejlin.}}
\author{Kazuhiko Sumiya\thanks{Corresponding author. Research Institute of Economy, Trade and Industry (RIETI), Tokyo, Japan; and Institute for Empirical Research in Organizational Economics, Waseda University, Tokyo, Japan. Email: \href{mailto:kazuhiko.sumiya@gmail.com}{kazuhiko.sumiya@gmail.com}} \and Jesper Bagger\thanks{School of Economics, University of Edinburgh, Edinburgh, United Kingdom; Dale T. Mortensen Centre, Department of Economics and Business Economics, Aarhus University, Aarhus, Denmark; and IZA, Bonn, Germany. Email: \href{mailto:jbagger@ed.ac.uk}{jbagger@ed.ac.uk}}}
\date{March 2025}

\maketitle

\begin{abstract}
\noindent This paper provides quasi-experimental evidence on how income taxes affect gross hourly wages, utilizing Danish administrative data and a tax reform that introduced joint taxation. Exploiting spousal income for identification, we present nonparametric, difference-in-differences graphical evidence among husbands. For low-income workers, taxes have negative and dynamic effects on wages; their wage elasticity with respect to net-of-marginal-tax rates is 0.4. For medium-income workers, the effects are smaller and insignificant. Wages respond to taxes through promotions or job-to-job transitions. Neither daily nor annual hours worked respond significantly; consequently, annual earnings respond to taxes primarily through hourly wages, rather than through labor supply.


\medskip


\end{abstract}

\clearpage

\section{Introduction}
Economists have long studied the distortionary and distributional effects of income taxes on labor market outcomes. While most empirical studies have focused on labor supply responses, there is limited evidence on how taxes affect gross hourly wages---another key component of earnings. Moreover, wages may respond to taxes through various channels. For example, some economists predict that higher tax rates have negative and dynamic effects on wages by reducing incentives to accumulate human capital or search for higher-paying jobs---two main drivers of wage growth \citep{Rubinstein2006}. Therefore, to better understand the policy implications, it is crucial to empirically identify both the effects of taxes on wages and the channels through which wages respond to taxes.

Given these motivations, this paper asks the following research questions: Do income taxes have positive or negative effects on gross hourly wages? Are these effects static or dynamic? Through what channels do wages respond to taxes? How do wage responses compare with the responses of annual earnings and labor supply?

To answer these questions compellingly, we provide quasi-experimental evidence utilizing Danish administrative data and a tax reform that came into effect in 1987. First, our dataset is a population-wide annual panel containing extensive information on individual income and worker/job characteristics. We select a sample of working married males. Next, we use the 1987 tax reform for our empirical strategy because it introduced joint taxation to a middle tax bracket, bringing large changes to the tax system faced by married couples. This unique institutional change allows us to exploit variation in spousal income and thus identify the following two groups located in a bottom tax bracket before the reform: one group with higher spousal income is pushed upward to the middle bracket by the reform, whereas the other group with lower spousal income remains in the bottom bracket due to joint taxation. We then compare the outcome dynamics of these two groups in a difference-in-differences (DID) design. Although \citet{Gruber2021} and \citet{Kleven2014b} also use this reform as a natural experiment, our approach is novel in its focus on the introduction of joint taxation.

We present clear nonparametric graphical evidence and regression results regarding the effects of taxes on various outcomes (see, for example, Figure \ref{fig:wage_low}). Table \ref{tab:elasticity} lists the main elasticity estimates with respect to net-of-marginal-tax rates, along with related results. Our findings are as follows.

First, taxes have heterogeneous effects on wages across income levels. Low-income workers respond to taxes negatively and dynamically. Higher marginal tax rates gradually depress wage growth over time; that is, distortion dynamically accumulates on wages, which we refer to as accumulating effects in this paper. Their elasticity of wages with respect to net-of-marginal-tax rates is 0.4. Their elasticity is relatively large because our DID design captures the dynamic and accumulating effects, rather than short-run effects attenuated by optimization frictions. Specifically, workers gradually change their behavior in response to taxes while overcoming optimization frictions. These gradual responses lead to the accumulating effects, which are reflected in their relatively large elasticity. By contrast, for medium-income workers, the effects of taxes on wages are smaller and statistically insignificant. Given the relatively large elasticity and novel accumulating effects, we move on to the details of low-income workers.

Second, we find that wages respond to taxes through promotions or job-to-job transitions. Workers facing higher marginal tax rates are less likely to be promoted to skilled or white-collar positions, arguably due to reduced incentives to accumulate human capital or work harder. Similarly, workers facing higher marginal tax rates are less likely to make job-to-job transitions, arguably due to reduced incentives to search for higher-paying jobs. These two dynamic channels can explain the negative and accumulating effects of taxes on wages. To the best of our knowledge, this paper provides the first quasi-experimental evidence on these channels. 

Third, we find that labor supply responses along the intensive margin, as measured by daily or annual hours worked, are statistically insignificant and smaller in magnitude than the wage responses. Consequently, the elasticity of annual earnings (estimated at 0.5) is primarily driven by that of hourly wages (estimated at 0.4), rather than by that of annual hours worked (estimated at 0.1 and insignificant).

Finally, the estimation results survive the following threats to identification: First, since spousal income serves as an instrumental variable in our DID design, we verify its exclusion restriction, which requires that spousal income affects outcome dynamics only through the treatment (i.e., tax rates). Specifically, in addition to confirming parallel pre-reform outcome dynamics, we conduct a placebo test and show parallel pre- and post-reform outcome dynamics between two placebo groups that differ exclusively in spousal income but face nearly identical tax rates (i.e., the absence of the treatment). Second, this placebo test also provides evidence that the introduction of joint taxation does not affect outcome dynamics through income reallocation between spouses, thereby validating our interpretation of wage responses as reflecting the effects of marginal tax rates. Third, although workers may become non-employed with missing wages, we show that the compositional changes of employed workers do not create the spurious negative effects of taxes on wages. Fourth, we confirm that individuals do not bunch at a tax bracket cutoff; otherwise, such bunching might bias estimates. Fifth, the estimation results are robust to modest changes to the definition of low-income workers.

This paper broadly contributes to the literature on public and labor economics. First, we offer novel implications for how taxes affect various labor market outcomes. Although \citet{Blomquist2010} and \citet{Martinez2021} also provide credible quasi-experimental evidence on negative wage responses, we found heterogeneous responses across income levels. Our findings on larger wage responses among low-income workers contrast with existing findings on larger taxable income responses among high-income earners \citep{Saez2012}, which has implications for progressive tax policies, such as the Earned Income Tax Credit in the United States. Furthermore, we found that wage responses occur through promotions or job-to-job transitions and are larger than labor supply responses, consequently accounting for most of annual earning responses---a series of new insights in the literature.

Second, this paper fills the gap between the micro and macro literature by estimating the accumulating effects. Broadly speaking, to study how income taxes affect labor market outcomes (e.g., wages), the micro literature estimates short-run elasticities using quasi-experimental methods, primarily DID designs \citep{Saez2012}. By contrast, the macro literature estimates long-run (steady-state) elasticities by calibrating dynamic models, such as learning-by-doing models \citep{Keane2012, Keane2015}, Ben-Porath models \citep{Heckman1998, Heckman1999}, or job search models \citep{Kreiner2015, Shephard2017}. Often, the micro literature reports smaller elasticities than the macro literature. We shed new light on this discrepancy as follows: By estimating the accumulating effects over a post-reform period of seven years in a credible DID design, we address optimization frictions that attenuate short-run elasticities estimated in the micro DID literature. As a result, we uncover structural elasticities that are conceptually akin to long-run elasticities estimated in the macro calibration literature \citep{Chetty2012, Chetty2011, Kleven2013, Kleven2023}. Therefore, the accumulating effects bridge these short- and long-run elasticities.

Finally, this paper provides motivations for welfare analysis and optimal taxation. Our findings lend empirical support to recent research on optimal policies with human capital accumulation or career effects \citep{Stantcheva2020}, or job search \citep{Chetty2008, Kroft2020}. Moreover, since we found larger wage responses than labor supply responses, it will be fruitful to extend optimal tax formulae based on the canonical labor supply model \citep{Piketty2013} by incorporating wage responses through alternative channels such as wage bargaining \citep{Piketty2014} or employer learning \citep{Craig2021}.

The paper proceeds as follows: Section \ref{sec:tax} describes the Danish income tax system and the 1987 tax reform. Section \ref{sec:emp_str} describes our empirical strategy. Section \ref{sec:data} describes the Danish administrative data. Section \ref{sec:DID} presents the estimation results. Section \ref{sec:conclusion} concludes. All tables and figures are included after the main text.

\section{The Danish income tax system and the 1987 tax reform}\label{sec:tax}
This section describes the Danish income tax system and the 1987 tax reform, focusing on their relevance to our empirical strategy. The key features of the tax system are a progressive structure with three tax brackets, and individual tax filing for married couples. The reform introduced joint taxation to a middle tax bracket, bringing large changes to the tax system faced by married couples.

\subsection{Income concepts}
In Denmark, income taxes are levied at the source with individual tax filing and are primarily based on three income concepts: labor income (LI), capital income (CI), and itemized deductions (D). For ease of exposition, we omit two income concepts of minor importance for our sample: personal income and stock income. These omitted income concepts are explained later in Appendix \ref{app:simulator}. However, we emphasize that we fully consider all income concepts when simulating tax liabilities for the empirical analysis in Section \ref{sec:emp_str}. Table \ref{tab:IncomeDef} lists the main items included in the three income concepts. Labor income is the main source of income for most individuals in our sample. Furthermore, their capital income is often negative due to interest payments on debt, such as mortgage loans.\footnote{Our description of the institutional settings relies on \citet{Kleven2014b}, the Danish Ministry of Finance (\url{https://fm.dk}), and the Danish Ministry of Taxation (\url{https://www.skm.dk}).}

\subsection{The pre-reform tax system}
The left panel of Table \ref{tab:1987reform} outlines the key features of the Danish income tax system in the final pre-reform year, 1986. Danish income taxes are divided into regional and national taxes. Regional taxes were levied on taxable income $\text{LI} + \text{CI} - \text{D}$ exceeding the cutoff of DKK 20,700 (DKK 1 in 1986 $=$ USD 0.3 in 2022). The regional tax rate is a flat rate that slightly varies across municipalities and counties; in 1986, the average regional tax rate was 28.0 percent, and the 90--10 percentile range was 3.7 percentage points.\footnote{For our sample period, 1981--1993, the administrative structure of Denmark comprised 275 municipalities (``kommuner'') and 14 counties (``amter'', a county that spans a set of municipalities). Both municipalities and counties levy income taxes on their residents. The regional tax rate is the sum of the municipal, county, and Church tax rates. The Church taxes are minuscule and are paid only by members of the Church of Denmark (``Folkekirken'').}

National taxes have a progressive structure with three tax brackets: bottom, middle, and top brackets. The tax base for each bracket was $\text{LI} + \text{CI} - \text{D}$, taxed with different cutoffs at different rates. These national taxes are cumulative, making the tax system progressive: taxable income $\text{LI} + \text{CI} - \text{D}$ exceeding the bottom bracket cutoff of DKK 23,200 was taxed at the bottom tax rate of 19.9 percent; $\text{LI} + \text{CI} - \text{D}$ exceeding DKK 113,400 was taxed at the middle tax rate of 14.4 percent; and $\text{LI} + \text{CI} - \text{D}$ exceeding DKK 186,100 was taxed at the top tax rate of 10.8 percent. Combined with the average regional tax rate of 28.0 percent, the marginal tax rates in 1986 were thus 47.9 percent in the bottom bracket, 62.3 percent in the middle bracket, and 73.1 percent in the top bracket. However, the top tax rate was adjusted downward to a marginal tax ceiling of 73.0 percent.

Crucially for our empirical strategy, taxation is based on individual tax filing for married couples. Married individuals file taxes individually and separately, with both spouses receiving equal tax treatment. Before the reform, even if married individuals were not liable for certain taxes, such as the middle taxes, and had unused allowances (calculated as the bracket cutoff DKK 113,400 minus their taxable income $\text{LI} + \text{CI} - \text{D}$), these unused allowances could not be transferred to their spouses. This point is highlighted in the left panel of Table \ref{tab:1987reform}, where the ``Joint" column is marked with ``No" indications.

\subsection{The 1987 tax reform}
The reform was legislated on March 18, 1986, and came into immediate and full effect on January 1, 1987. We describe the background of the reform in Appendix \ref{app:tax}.

The right panel of Table \ref{tab:1987reform} outlines the key features of the Danish income tax system in 1987. There were only minor changes in regional taxes. In national taxes, there were three major changes. First, the bottom and top tax rates were increased, whereas the middle tax rate was decreased; as a result, combined with the average regional tax rate of 29.0 percent, the marginal tax rates in 1987 were 51.0 percent in the bottom bracket, 57.0 percent in the middle bracket, and 69.0 percent in the top bracket. 

Second, the tax base for the middle bracket was changed to labor income plus positive capital income ($\text{LI} + [\text{CI}>0]$), and the tax base for the top bracket was changed to labor income plus capital income exceeding DKK 60,000 ($\text{LI} + [\text{CI}>60\text{k}]$). In addition, the three bracket cutoffs were all increased by more than a statutory inflation adjustment of 2.0 percent. As a result, some individuals were mechanically pushed to other brackets. 

Finally and most crucially for our empirical strategy, the reform introduced joint taxation to the middle bracket. If married individuals are not liable for the middle taxes and have unused allowances (calculated as the bracket cutoff DKK 130,000 minus their taxable income $\text{LI} + [\text{CI}>0]$), these unused allowances can be transferred to their spouses. To elaborate on this institutional change, consider the following example: suppose, in 1987, a husband (our sample) has taxable income for the middle bracket equal to DKK 150,000 (i.e., $\text{LI} + [\text{CI}>0] = 150,000$), and his wife has taxable income equal to DKK 100,000 (i.e., $\text{LI}^\text{w} + [\text{CI}^\text{w}>0] = 100,000$, with the superscript w denoting ``wife"). If the joint taxation is not in place (like in the pre-reform tax system), he is liable for the middle taxes given the bracket cutoff of DKK 130,000, but she is not. Note that she has unused allowances equal to $130,000 - 100,000 = 30,000$. Under the post-reform tax system, her unused allowances can be transferred to him; therefore, his taxable income is re-calculated as $150,000 - 30,000 = 120,000$. Then, he is no longer liable for the middle taxes. Despite this transfer scheme, married individuals continue to file taxes individually and separately; admittedly, using the words ``joint taxation" might be a slight abuse of terminology.

We utilize this institutional change as a novel natural experiment. Specifically, in the next section, we identify two groups of individuals with similar income patterns (LI, CI, D, $\text{CI}^\text{w}$, $\text{D}^\text{w}$) who were in the bottom bracket before the reform. One group with higher wives' labor income ($\text{LI}^\text{w}$) is mechanically pushed upward to the middle bracket by the reform, whereas the other group with lower wives' labor income ($\text{LI}^\text{w}$) remains in the bottom bracket due to the joint taxation. These bracket movements form the basis for the DID design employed in this paper.

\section{Empirical strategy}\label{sec:emp_str}
This section explains how we select a sample, define treated and control individuals in our DID design, and address endogeneity caused by the correlation between treatment assignment and pre-reform income. Our empirical strategy leverages the joint taxation introduced to the middle bracket and cross-sectional variation in wives' labor income ($\text{LI}^\text{w}$).

\subsection{Sample selection}
We select a sample period 1981--1993 for two reasons. First, a reliable measure of gross hourly wages is available from 1981 \citep{Lund2016}. Second, the subsequent tax reform after the one in 1987 occurred in 1994. The year 1986 serves as a baseline pre-reform year for this paper. Although we detail our data in Section \ref{sec:data}, it suffices to mention here that we have access to annual panel data on individual income (LI, CI, D, $\text{LI}^\text{w}$, $\text{CI}^\text{w}$, $\text{D}^\text{w}$), demographic characteristics, and outcomes (e.g., wages) for our sample period 1981--1993.

We exclude females from our sample to facilitate identification. First, \citet{Chetty2011} find that in Denmark, married females tend to bunch at bracket cutoffs. Since our DID design is based on bracket movements, such bunching complicates identification. Second, \citet{Kleven2019} find that in Denmark, more than 10 percent of female workers (as opposed to almost zero percent of male workers) exit the labor market after having a child. These labor market exits further complicate the identification of wage responses to taxes among employed workers. 

Consequently, we select a sample of prime-age working males whose wives had positive labor income in 1986. Specifically, we select males who, in 1986, were (i) younger than 50 years old and (ii) employed on the 28th of November---our definition of employment (to be explained in Section \ref{sec:data}). These two restrictions identify males strongly attached to the labor market, facilitating the identification of wage responses to taxes among employed workers. We further impose the following two restrictions on the sample: in 1986, (iii) they were married, and (iv) their wives had (strictly) positive labor income. We impose the third restriction to exploit the joint taxation introduced by the reform. We impose the fourth restriction to make treated and control individuals similar by excluding non-working wives; thus, we exploit cross-sectional variation in labor income earned by working wives, as shown in the following subsections.

Hence, our sample consists of males who satisfy the four restrictions related to the baseline pre-reform year 1986 and are tracked both backward until 1981 and forward until 1993. We have small sample attrition due to reasons such as death or emigration, with a 2.2\% attrition rate in 1993. Based on this sample, we define treated and control individuals below.

\subsection{Treated and control individuals}
Since our DID design focuses on movements between the bottom and middle brackets, we first identify the tax brackets in which individuals are located using a tax simulator. The simulator was originally developed by \citet{Kleven2014b} and \citet{Bagger2018}, and encodes the details of the Danish income tax system for each year. It identifies a bracket location for each individual by taking as its main input his and spousal income (LI, CI, D, $\text{LI}^\text{w}$, $\text{CI}^\text{w}$, $\text{D}^\text{w}$); see Appendix \ref{app:simulator} for an overview of the simulator. Let $\text{B}^{86}(z_{i86})$ denote that individual $i$ with 1986 income $z_{i86}:=\{\text{LI}_{i86}, \text{CI}_{i86}, \text{D}_{i86}, \text{LI}^\text{w}_{i86}, \text{CI}^\text{w}_{i86}, \text{D}^\text{w}_{i86}\}$ is in the bottom bracket under the 1986 tax system. More precisely, $\text{B}^{86}(z_{i86})$ means that he is liable for the bottom taxes but neither the middle nor top taxes.

To define treated and control individuals based on bracket movements, we next consider the following counterfactual bracket location: let $\widetilde{\text{B}}^{87}(z_{i86})$ denote that individual $i$ with 1986 income $z_{i86}$ (rather than 1987 income $z_{i87}$) is in the bottom bracket if hypothetically facing the inflation-adjusted 1987 tax system. We adjust all monetary values related to the 1987 tax system (e.g., bracket cutoffs) to the 1986 price level. Analogously, let $\widetilde{\text{M}}^{87}(z_{i86})$ denote that individual $i$ with 1986 income $z_{i86}$ is in the middle bracket under the inflation-adjusted 1987 tax system. More precisely, $\widetilde{\text{M}}^{87}(z_{i86})$ means that he is liable for both the bottom and middle taxes but not the top taxes. We identify individuals in either $\widetilde{\text{B}}^{87}(z_{i86})$ or $\widetilde{\text{M}}^{87}(z_{i86})$ using the tax simulator.

By combining the actual and counterfactual bracket locations, we define treated and control individuals as follows:
\begin{align}\label{eq:treated}
  \text{Treated: } & \ \text{B}^{86}(z_{i86}) \ \ \text{and} \ \ \widetilde{\text{M}}^{87}(z_{i86}) \notag \\
  \text{Control: } & \ \text{B}^{86}(z_{i86}) \ \ \text{and} \ \ \widetilde{\text{B}}^{87}(z_{i86}).
\end{align}
The treated individuals were in the bottom bracket in the baseline pre-reform year 1986 but are in the middle bracket under the 1987 tax system. $\widetilde{\text{M}}^{87}(z_{i86})$ is independent of behavioral responses to the reform because their income is fixed at the pre-reform 1986 level; therefore, their movement from the bottom bracket to the middle bracket is mechanically created by the reform. By contrast, the control individuals remain in the bottom bracket in the absence of behavioral responses to the reform, i.e., $\widetilde{\text{B}}^{87}(z_{i86})$. Hence, the treatment is being mechanically pushed upward to the middle bracket, as \citet{Saez2003} similarly exploits bracket creep caused by high inflation.\footnote{One might consider another bracket movement that exploits the joint taxation introduced to the middle bracket: treated individuals are pushed downward from the middle bracket to the bottom bracket, whereas control individuals remain in the middle bracket. In this case, however, we found that these individuals were not similar in pre-reform covariates or trends. By contrast, we show in the next subsection that the treated and control individuals defined by the treatment assignment \eqref{eq:treated} are similar, which lends credibility to our DID design.}

\subsection{Endogeneity caused by pre-reform income}
The treatment assignment \eqref{eq:treated} correlates with pre-reform income $z_{i86}$, which potentially causes endogeneity: if the treated individuals had significantly higher $z_{i86}$ than the control individuals, their outcomes (e.g., wages) would evolve differently even without any treatment. One thus needs to control for $z_{i86}$; however, controlling for every component of $z_{i86}=\{\text{LI}_{i86}, \text{CI}_{i86}, \text{D}_{i86}, \text{LI}^\text{w}_{i86}, \text{CI}^\text{w}_{i86}, \text{D}^\text{w}_{i86}\}$ loses variation for identification---a challenge recognized in the literature on the elasticity of taxable income \citep{Saez2012}. We show below that nonparametrically controlling for labor income $\text{LI}_{i86}$ also balances own and spousal capital income and deductions ($\text{CI}_{i86}$, $\text{D}_{i86}$, $\text{CI}^\text{w}_{i86}$, $\text{D}^\text{w}_{i86}$); as a result, our empirical strategy leverages variation in wives' labor income $\text{LI}_{i86}^\text{w}$ as a source of identification.

Figure \ref{fig:LI_density} presents the distributions of pre-reform labor income $\text{LI}_{i86}$. The left panel plots the kernel density estimates of $\text{LI}_{i86}$ by treatment status. The two distributions are similar and sufficiently overlap. The large overlap is thanks to the joint taxation and variation in wives' income. We elaborate on this point in Appendix \ref{app:single} by demonstrating that among single males, the distributions of $\text{LI}_{i86}$ do not sufficiently overlap between treated and control individuals. 

Returning to married males (our sample) in Figure \ref{fig:LI_density}, we plot in the right panel the histogram of $\text{LI}_{i86}$ with a bin width of DKK 20,000 among the treated and control individuals (on the left axis). We also plot the fractions of the treated individuals within each bin (on the right axis). These fractions can be regarded as propensity scores given $\text{LI}_{i86}$. Two observations are worth highlighting. First, there is limited overlap in the two leftmost bins. Therefore, we discard individuals in these bins---a decision partially informed by a rule of thumb proposed by \citet{Crump2009} to discard individuals with estimated propensity scores outside the range $[0.1, 0.9]$. Second, the propensity scores vary within the domain $[120\text{k}, 180\text{k}]$ with large sample sizes but remain constant within the domain $[180\text{k}, 280\text{k}]$ with small sample sizes.

Based on these observations, we define two income groups as follows: low-income ($120,000 \leq \text{LI}_{i86} < 160,000$) and medium-income ($160,000 \leq \text{LI}_{i86} < 280,000$) groups. The low-income group has a narrower domain of $\text{LI}_{i86}$ that limits the impact of the varying propensity scores while still retaining a sufficient sample size. By selecting this narrow domain, we nonparametrically control for $\text{LI}_{i86}$, thus expecting balanced $\text{LI}_{i86}$ between the treated and control individuals. By contrast, the medium-income group has a wider domain that ensures a sufficient sample size, with the constant propensity scores contributing to balanced $\text{LI}_{i86}$. Therefore, the way we construct these two groups is similar to stratification based on the propensity scores. We analyze each group separately and later check robustness to modest changes to the group definition.

Next, we present summary statistics of pre-reform covariates to assess the similarity between the treated and control individuals. Table \ref{tab:Covariate} lists the means of covariates in 1986 by treatment status for each income group. Standard deviations are in parentheses. Notice that, by construction, the two income groups have sufficient and roughly identical sample sizes. Following a common practice in the matching literature \citep{Imbens2015, Imbens2015b, Stuart2010}, we assess the similarity using a normalized difference (abbreviated as ``N. d." in the table). It is defined as $\frac{\overline{X}_T - \overline{X}_C}{\sqrt{(S_T^2 + S_C^2) / 2}}$ for each covariate, where $\overline{X}_T$ and $\overline{X}_C$ are the means among the treated and control individuals respectively, and $S_T$ and $S_C$ are the corresponding standard deviations. Importantly, the normalized difference is independent of sample sizes in expectation, thus offering a key advantage over the $t$-statistic for testing the null hypothesis of no differences in population means between the treated and control individuals. As a general guideline, a covariate is considered as balanced between the treated and control individuals if its normalized difference is less than 0.25. 

Table \ref{tab:Covariate} shows the same patterns for the two income groups. First, by construction, labor income $\text{LI}_{i86}$ is balanced between the treated and control individuals. Although the normalized difference within the low-income group is relatively large (0.65), the raw difference in mean $\text{LI}_{i86}$ is modest (DKK 6,000) and only slightly larger than that within the medium-income group (DKK 4,000). Note that DKK 1 in 1986 equals USD 0.3 in 2022. Therefore, we consider $\text{LI}_{i86}$ as balanced within the low-income group---a conclusion later supported by the insensitivity of estimation results to modest changes to the group definition ($120,000 \leq \text{LI}_{i86} < 160,000$).\footnote{The narrow domain of $\text{LI}_{i86}$ for the low-income group limits the impact of the varying propensity scores (as displayed in the right panel of Figure \ref{fig:LI_density}) and thus contributes to the modest difference in mean $\text{LI}_{i86}$ (DKK 6,000). At the same time, it also contributes to small standard deviations (DKK 8,000--11,000), thereby inflating the normalized difference to 0.65.}

Let us move on to the other covariates listed in Table \ref{tab:Covariate}. Once $\text{LI}_{i86}$ is controlled for, the following covariates are also balanced with normalized differences less than 0.25: worker/job characteristics (age, ..., private-sector jobs), and own and spousal capital income and deductions ($\text{CI}_{i86}$, $\text{D}_{i86}$, $\text{CI}_{i86}^\text{w}$, $\text{D}_{i86}^\text{w}$). These balanced covariates lend credibility to our estimation results. As a result, the main difference between the treated and control individuals lies in wives' labor income $\text{LI}_{i86}^\text{w}$. The treated individuals have higher $\text{LI}_{i86}^\text{w}$ and thus are pushed upward to the middle bracket by the reform. By contrast, the control individuals have lower $\text{LI}_{i86}^\text{w}$ and thus remain in the bottom bracket due to the joint taxation. Although using variation in spousal income as a source of identification is not new in the literature, e.g., \citet{Eissa1995, Eissa1996}, we leverage it with the joint taxation introduced by the reform.

Finally, we clarify our identification assumption. The DID design assumes that, in the absence of the treatment, the average outcome among the treated individuals evolves in parallel with that among the control individuals \citep{Lechner2010}. Since these individuals differ exclusively in wives' labor income $\text{LI}_{i86}^\text{w}$, this assumption essentially states that $\text{LI}_{i86}^\text{w}$ serves as an instrumental variable: it affects the treatment (i.e., bracket locations and, thus, tax rates) and, only through the treatment, affects outcome dynamics. Although we confirm the first requirement (i.e., relevance) in Section \ref{sec:DID}, the second requirement (i.e., exclusion restriction) is not directly testable; however, parallel pre-reform outcome dynamics provide supporting evidence. Moreover, in Section \ref{sec:DID}, we conduct a placebo test and show parallel pre- and post-reform outcome dynamics between placebo-treated and placebo-control individuals who differ exclusively in $\text{LI}_{i86}^\text{w}$ but face nearly identical tax rates (i.e., the absence of the treatment).

\section{Danish administrative data}\label{sec:data}
We utilize population-wide Danish administrative data constructed from three register-based sources: tax return data, IDA (an acronym for ``Integreret Database for Arbejdsmarkedsforskning" translated as the Integrated Database for Labor Market Research), and job spell data. These data are maintained by Statistics Denmark, cover all legal residents in Denmark aged 15--74 (on the 31st of December each year) since 1980, and share a common individual ID.

\paragraph{Tax return data.}
The tax return data are effectively identical to the annual panel data used by \citet{Kleven2014b}, and contain variables such as marital status and precisely measured individual income (LI, CI, D, $\text{LI}^\text{w}$, $\text{CI}^\text{w}$, $\text{D}^\text{w}$). We use the tax return data as, among other things, inputs to the tax simulator to simulate the bracket locations (used in Section \ref{sec:emp_str} to define the treated and control individuals) and effective marginal tax rates (used in Section \ref{sec:DID} to compute elasticities); see Appendix \ref{app:simulator} for details on the tax return data in the overview of the tax simulator.

\paragraph{IDA.}
IDA is an annual panel constructed from several registers (e.g., social security and tax registers), and contains extensive information on workers and jobs, such as gross hourly wages and worker/job characteristics listed in Table \ref{tab:Covariate}. In the data, we observe employment status on the 28th of November each year. Employment is thus defined as holding a primary job on this specific date, referred to as a November job hereafter in this paper. IDA contains gross hourly wages for a November job each year, which serve as our key outcome variable; see Appendix \ref{app:variable} for details on the computation of gross hourly wages. Finally, note that we distinguish \textit{hourly} wages for a November job from \textit{annual} labor income (LI).

\paragraph{Job spell data.}
The job spell data are constructed from employer-reported income tax reports, cover all primary job spells since 1985, and contain (i) the start and end dates of each spell and (ii) hours worked in each spell each year. The unit of observation is thus person-spell-year, as opposed to person-year in the tax return data and IDA. To link these three data sources, we construct an annual panel of November jobs from the job spell data by extracting job spells ongoing on the 28th of November each year. In this ``November-job" annual panel, the unit of observation is person-year. Using the variables (i) and (ii), we then compute daily hours worked for a November job each year, which serve as our second outcome variable; see Appendix \ref{app:variable} for details on the computation of daily hours worked.

\paragraph{Data construction.}
We link the tax return data, IDA, and the ``November-job" data using the common individual ID each year over the sample period 1981--1993. Although wages and hours are missing among non-employed individuals observed in the tax return data and IDA, we retain all observations regardless of employment status. The constructed dataset is an annual panel covering 1981--1993 and contains the following variables: wages, hours, worker/job characteristics, individual income, bracket locations, and effective marginal tax rates. The availability of both gross hourly wages and daily hours worked, together with the tax return data, is a novel and advantageous feature of our dataset.\footnote{In addition to these two outcome variables, we also utilize gross annual earnings, annual hours worked, and variables regarding promotions and job-to-job transitions, all of which are described in the corresponding analyses below.}

\section{Estimation results}\label{sec:DID}
This section presents nonparametric graphical evidence, regression results, and implied elasticities regarding the effects of taxes on various outcomes. We first study wage responses and subsequently show that these results survive all threats to identification. To analyze the channels through which wages respond to taxes, we next study promotions and job-to-job transitions. Finally, we compare the responses of hourly wages, annual earnings, and labor supply.

\subsection{Wage responses by the low-income group}
\paragraph{Graphical evidence.}
Figure \ref{fig:wage_low} presents wage responses by the low-income group ($120,000 \leq \text{LI}_{i86} < 160,000$). Outcome $Y_{it}$ is the $\log$ of real gross hourly wages for a November job that individual $i$ holds in year $t$. The left panel plots $\overline{Y_{t}} - \overline{Y_{86}}$ for $t = 81, ..., 93$ by treatment status, where $\overline{Y_{t}}$ denotes mean $Y_{it}$ over $i$. Before the reform, the wage dynamics are parallel, which supports our identification assumption. After the reform, the treated individuals, who are pushed upward from the bottom bracket to the middle bracket, experience lower wage growth than the control individuals, who remain in the bottom bracket. Furthermore, the wage dynamics gradually diverge: the wage growth rate among the treated individuals is lower than that among the control individuals by a small margin from 1986 to 1987, but by approximately 1.5 percentage points from 1986 to 1993---a sizable difference given that their overall wage growth rate is approximately 15 percent from 1986 to 1993. The left panel provides compelling nonparametric graphical evidence regarding the negative and accumulating effects of taxes on wages.

\paragraph{Regression analysis.}
To facilitate inference, we turn to regression analysis and estimate the following two-way fixed effect model for individual $i$ and year $t = 81, ..., 93$:
\begin{equation}\label{eq:DID}
    Y_{it} = \alpha_i + \sum_{j \ne 86} \alpha_{j} \cdot Year_{t = j} + \sum_{j \ne 86} \beta_{j} \cdot Year_{t = j} \cdot Treated_i + u_{it}.
\end{equation}
This specification is standard in the literature on DID and event studies \citep{Miller2023}. Outcome $Y_{it}$ remains as defined above: the $\log$ of real gross hourly wages for a November job that individual $i$ holds in year $t$. $\alpha_i$ is an individual fixed effect. $\alpha_j$ is a year fixed effect, $Year_{t = j}$ is a dummy variable that equals one if year $t$ equals $j$, and $t = 86$ serves as the excluded reference year. $Treated_i$ is a dummy variable that equals one if individual $i$ is treated. $\beta_t$ are the parameters of interest and measure differences in wage dynamics between the treated and control individuals before the reform ($\beta_{81}$, ..., $\beta_{85}$) and after the reform ($\beta_{87}$, ..., $\beta_{93}$). $u_{it}$ is an error term. Standard errors are clustered at the individual level \citep{Bertrand2004}.

The right panel of Figure \ref{fig:wage_low} plots the point estimates of $\beta_t$ for $t=81, ..., 93$ with their 95\% confidence intervals. The regression results align with the graphical evidence in the left panel. First, reassuringly, none of the pre-reform effects ($\widehat{\beta_{81}}$, ..., $\widehat{\beta_{85}}$) statistically differ from zero, which suggests that the parallel trends assumption is plausible. Next, the post-reform effect for 1987 ($\widehat{\beta_{87}}$) is statistically insignificant, implying sluggish responses due to optimization frictions. The subsequent post-reform effects ($\widehat{\beta_{88}}$, ..., $\widehat{\beta_{93}}$) are all negative and significant. Furthermore, the point estimates tend to decrease over time, approximately from zero to $-$0.015. These results clearly show the dynamic and accumulating effects of taxes on wages.

\begin{figure}[!htbp]\centering
\caption{Bracket locations for the low-income group}
\label{fig:bracket_low}
\medskip
\includegraphics[scale=0.6]{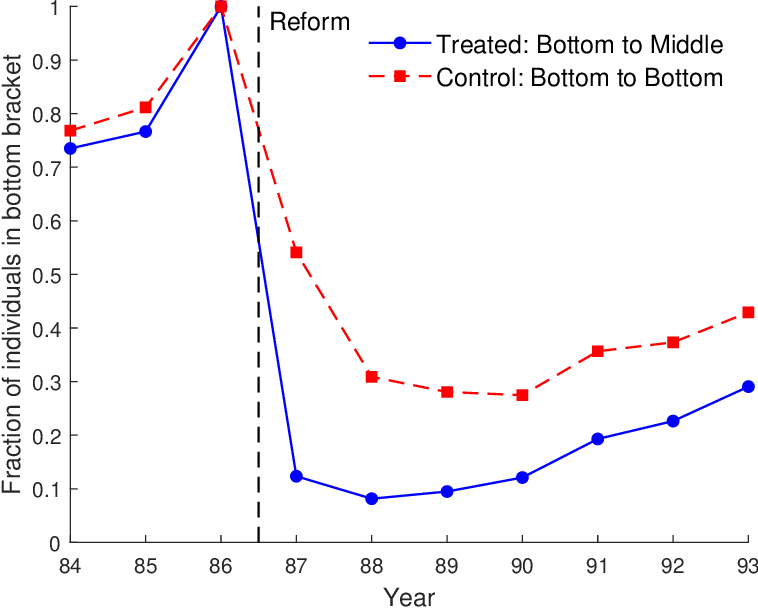}\hfill\includegraphics[scale=0.6]{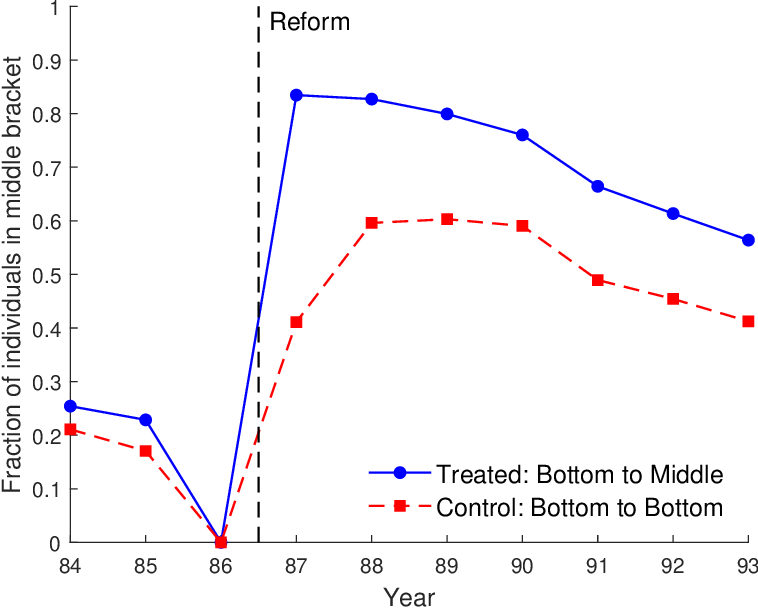}
\medskip
\caption*{\footnotesize Notes: The figure presents bracket locations for the low-income group ($120,000 \leq \text{LI}_{i86} < 160,000$). The left (right) panel plots the fractions of individuals located in the bottom (middle, respectively) bracket by treatment status. Bracket locations from 1981 to 1983 are missing due to data limitations. The sample is males who, in 1986, were (i) younger than 50 years old and (ii) employed on the 28th of November. Furthermore, in 1986, (iii) they were married, and (iv) their wives had (strictly) positive labor income. The treated and control individuals are defined by the treatment assignment \eqref{eq:treated}.}
\end{figure}

\paragraph{Bracket locations.}
Next, we examine bracket locations in Figure \ref{fig:bracket_low}. The left (right) panel plots the fractions of individuals located in the bottom (middle, respectively) bracket by treatment status. Bracket locations from 1981 to 1983 are missing due to data limitations. The treated and control individuals were in similar brackets on average between 1984 and 1985, and in the bottom bracket in 1986, as defined by the treatment assignment \eqref{eq:treated}. Although the treated individuals are pushed upward to the middle bracket by the reform, their bracket movement is purely mechanical: in the absence of behavioral responses to the reform, they are in the middle bracket under the 1987 tax system, i.e., $\widetilde{\text{M}}^{87}(z_{i86})$ in the treatment assignment \eqref{eq:treated}. Given that their income generally changes in 1987, i.e., $z_{i86} \neq z_{i87}$, their actual brackets in 1987 may deviate from the middle bracket. Despite this non-compliance with the assigned tax bracket, the treated individuals are 20--40 percentage points more likely to be in the middle bracket from 1987 onward. Finally, Figure \ref{fig:bracket_low_app} in Appendix \ref{app:fig} shows that small fractions of individuals are located in the top bracket (in the left panel) or in none of the three brackets, i.e., not liable for national taxes (in the right panel), without noticeable differences between the treated and control individuals.

\paragraph{Elasticity.}
Due to non-compliance with the assigned tax brackets, the estimated DID coefficients $\widehat{\beta_t}$ in Equation \eqref{eq:DID} should be interpreted as intention-to-treat (ITT) effects. On the other hand, treatment-on-the-treated (TOT) effects, which scale up ITT effects to account for non-compliance, are also of policy interest. Furthermore, to put the estimation results into context, we need an elasticity with respect to net-of-tax rates. Therefore, we convert the year-by-year ITT effects $\widehat{\beta_t}$ for the post-reform period into a TOT elasticity. Inspired by \citet{Gruber2021} and \citet{Jakobsen2020}, we compute this elasticity in two steps: first, we convert the year-by-year ITT effects $\widehat{\beta_t}$ into a TOT effect using the treatment assignment $Treated_i$ as an instrumental variable (IV), and then we convert this TOT effect into a TOT elasticity using mechanical changes in net-of-tax rates.\footnote{Alternatively, one could use wives' labor income in 1986 ($\text{LI}_{i86}^\text{w}$), which is our source of identification, as a continuous IV. In such cases, researchers often transform or discretize continuous IVs \citep{Angrist2009}. Due to the lack of general guidelines on these procedures, we opt to use the binary IV $Treated_i$.}

In the first step, we compute a TOT effect by estimating the following two-way fixed effect model for individual $i$ and year $t = 84, ..., 93$:
\begin{equation}\label{eq:TOT}
    Y_{it} = \alpha_i + \sum_{j \ne 86} \alpha_{j} \cdot Year_{t = j} + \beta^\text{TOT} \cdot Post_t \cdot M_{it} + u_{it}.
\end{equation}
$Post_t$ is a dummy variable that equals one if year $t$ is in the post-reform period, i.e., $t \geq 87$. $M_{it}$ is a dummy variable that equals one if individual $i$ is located in the middle or top brackets in year $t$, and zero if located in the bottom or none of the three brackets. As already shown in Figure \ref{fig:bracket_low_app} of Appendix \ref{app:fig}, only small fractions of individuals are in the top or none of the three brackets; thus, $M_{it}$ effectively indicates whether individual $i$ is in the middle or bottom brackets in year $t$. $\beta^\text{TOT}$ measures the effect of $M_{it}$ averaged over the post-reform period. Standard errors are clustered at the individual level. Notice that in this regression, year $t$ starts from 84, rather than 81, because bracket locations from 1981 to 1983 are missing, which does not pose a concern given our focus on the post-reform period.

Since $Post_t \cdot M_{it}$ in Equation \eqref{eq:TOT} is endogenous due to reverse causality, we instrument it with $Post_t \cdot Treated_i$. Three points are worth highlighting. First, in the reduced form of this IV estimation, the coefficient on $Post_t \cdot Treated_i$ (with $Y_{it}$ on the left-hand side) corresponds to the average of the year-by-year ITT effects $\beta_t$ in Equation \eqref{eq:DID} over the post-reform period. Second, in the first stage of this IV estimation, the coefficient on $Post_t \cdot Treated_i$ (with $Post_t \cdot M_{it}$ on the left-hand side) captures the persistent differences in post-reform bracket locations between the treated and control individuals illustrated in Figure \ref{fig:bracket_low}. Third, the exclusion restriction is supported by the parallel pre-reform trends found in Figure \ref{fig:wage_low}, along with almost no differences in pre-reform bracket locations found in Figure \ref{fig:bracket_low}.

In the second step, we compute the elasticity of outcome $Y$ (e.g., wages) with respect to net-of-tax rates as follows:
\begin{equation}\label{eq:elasticity}
    \widehat{\epsilon} := \frac{\widehat{\beta^\text{TOT}}}{\text{avg}[ \ \Delta \log(1-\tau_{i}^\text{Mech}) \ | \ \text{Treated} \ ] - \text{avg}[ \ \Delta \log(1-\tau_{i}^\text{Mech}) \ | \ \text{Control} \ ]}.
\end{equation}
The two terms in the denominator represent mechanical changes in net-of-tax rates among the treated and control individuals, respectively. Specifically, $\text{avg}[ \ \cdot \ | \ \text{Treated} \ ]$ denotes an average over the treated $i$, and the mechanical changes are defined as
\begin{equation*}
    \Delta \log(1-\tau_{i}^\text{Mech}) := \log(1-\widetilde{\tau}^{87}(z_{i86})) - \log(1-\tau^{86}(z_{i86})).
\end{equation*}
$\tau^{86}(z_{i86})$ and $\widetilde{\tau}^{87}(z_{i86})$ denote tax rates that individual $i$ with 1986 income $z_{i86}$ faces under the 1986 tax system and under the inflation-adjusted 1987 tax system, respectively. This definition of the mechanical changes aligns with the treatment assignment \eqref{eq:treated} and appropriately captures the reform-induced variation in tax rates. 

We compute $\tau^{86}(z_{i86})$ as an \textit{effective} marginal tax rate on labor income (LI) using the tax simulator, which encodes the details of the Danish income tax system. Specifically, given the main input $z_{i86}=\{\text{LI}_{i86}, \text{CI}_{i86}, \text{D}_{i86}, \text{LI}^\text{w}_{i86}, \text{CI}^\text{w}_{i86}, \text{D}^\text{w}_{i86}\}$, it is computed as
\begin{equation*}
  \tau^{86}(z_{i86}) := \frac{T^{86}(\text{LI}_{i86} + 100, \text{CI}_{i86}, \text{D}_{i86}, \text{LI}_{i86}^\text{w}, \text{CI}_{i86}^\text{w}, \text{D}_{i86}^\text{w}) - T^{86}(\text{LI}_{i86}, \text{CI}_{i86}, \text{D}_{i86}, \text{LI}_{i86}^\text{w}, \text{CI}_{i86}^\text{w}, \text{D}_{i86}^\text{w})} {100},
\end{equation*}
where $T^{86}(\cdot)$ denotes simulated tax liabilities under the 1986 tax system (DKK 100 in 1986 $=$ USD 30 in 2022). $\widetilde{\tau}^{87}(z_{i86})$ is computed analogously.

Table \ref{tab:elasticity} lists the main elasticity estimates, along with related results. Here, we review findings on gross hourly wages for the low-income group. The first stage of the IV estimation is strong: the coefficient on the IV is estimated to be 0.229 and highly significant, with an $F$-statistic greater than 104.7. Note that \citet{Lee2022} show that a first-stage $F$-statistic greater than 104.7 ensures a correct size of 5\% for a two-sided $t$-test for a 2SLS coefficient in broad settings of just-identified, single IV cases. Since the first-stage results are almost identical across outcomes, we report them only for gross hourly wages. The IV estimation, specified by Equation \eqref{eq:TOT}, yields a TOT effect $\widehat{\beta^\text{TOT}}$ of $-$0.044, which represents the average of the year-by-year ITT effects $\widehat{\beta_t}$ over the post-reform period, scaled up by the first-stage effect. At the bottom of Table \ref{tab:elasticity}, we report the mechanical changes in net-of-tax rates by treatment status. Although not reported in the table, the standard deviations of $\Delta \log(1-\tau_{i}^\text{Mech})$ are only approximately 10$^{-4}$; thus, we treat these mechanical changes as constants. Finally, we obtain an elasticity $\widehat{\epsilon}$ of 0.369, with a standard error ($\widehat{\text{s.e.}}$) of 0.065 computed using the delta method.

\subsection{Wage responses by the medium-income group}
\begin{figure}[!htbp]\centering
\caption{Wage responses by the medium-income group}
\label{fig:wage_medium}
\medskip
\includegraphics[scale=0.6]{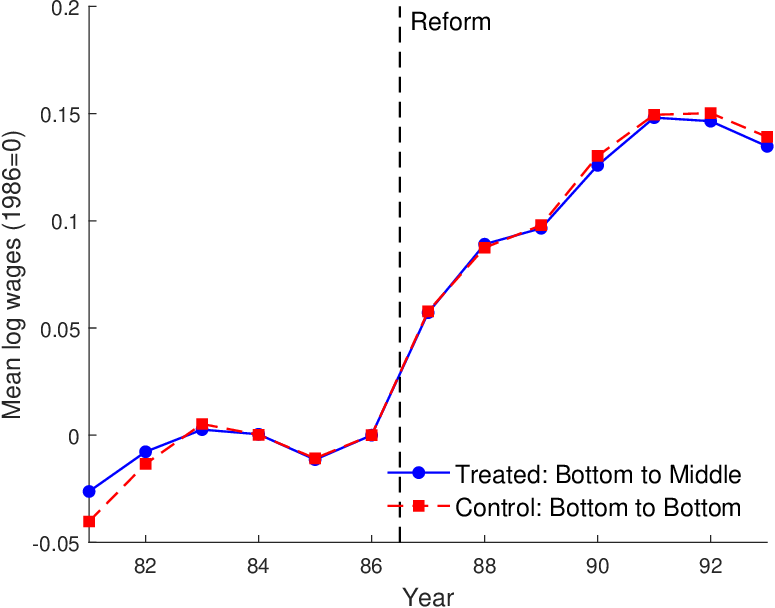}\hfill\includegraphics[scale=0.6]{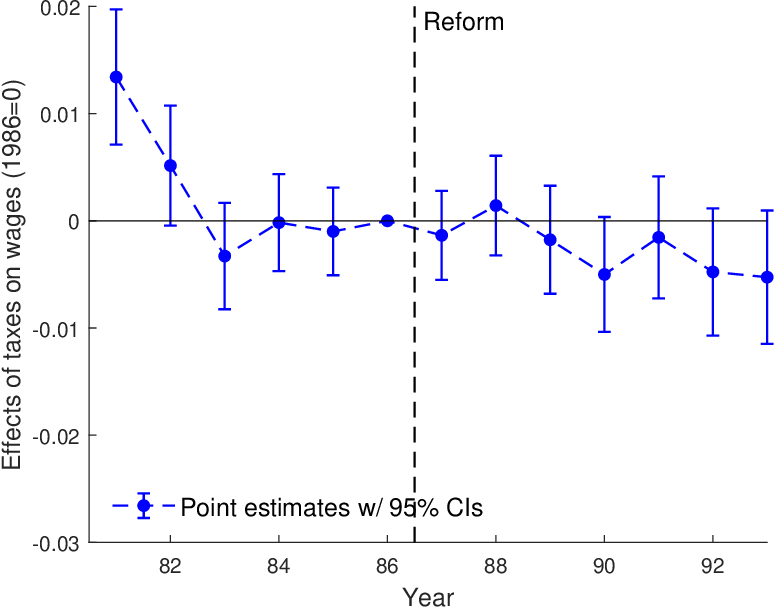}
\medskip
\caption*{\footnotesize Notes: The figure presents wage responses by the medium-income group ($160,000 \leq \text{LI}_{i86} < 280,000$). Outcome $Y_{it}$ is the $\log$ of real gross hourly wages for a November job that individual $i$ holds in year $t$. The left panel plots $\overline{Y_{t}} - \overline{Y_{86}}$ for $t = 81, ..., 93$ by treatment status, where $\overline{Y_{t}}$ denotes mean $Y_{it}$ over $i$. The sample is males who, in 1986, were (i) younger than 50 years old and (ii) employed on the 28th of November. Furthermore, in 1986, (iii) they were married, and (iv) their wives had (strictly) positive labor income. The treated and control individuals are defined by the treatment assignment \eqref{eq:treated}. The right panel plots the point estimates of $\beta_t$ for $t=81, ..., 93$ with their 95\% confidence intervals from the two-way fixed effect model specified by Equation \eqref{eq:DID}. Standard errors are clustered at the individual level.}
\end{figure}

\paragraph{Results.}
We repeat the analysis for the medium-income group ($160,000 \leq \text{LI}_{i86} < 280,000$) by providing graphical evidence, DID coefficients $\widehat{\beta_t}$, and an implied elasticity $\widehat{\epsilon}$. Figure \ref{fig:wage_medium} presents the graphical evidence and DID coefficients $\widehat{\beta_t}$. Note that the figures related to wage responses, such as Figures \ref{fig:wage_low} and \ref{fig:wage_medium}, use the same y-axis scale for ease of comparison. The left panel shows that the wage dynamics are almost parallel before and after the reform. Although $\widehat{\beta_{81}}$ statistically differs from zero, the subsequent pre-reform effects ($\widehat{\beta_{82}}$, ..., $\widehat{\beta_{85}}$) do not. In addition, the post-reform effects ($\widehat{\beta_{87}}$, ..., $\widehat{\beta_{93}}$) are all insignificant. As a result, the implied elasticity $\widehat{\epsilon}$ is 0.112 ($\widehat{\text{s.e.}} = 0.089$), which is also insignificant and smaller than that of the low-income group (refer to Table \ref{tab:elasticity} for a summary including strong first-stage results).

\paragraph{Summing up.}
We found heterogeneous wage responses across income levels. The low-income group responds to taxes negatively and dynamically, with an elasticity of 0.4. Their elasticity is relatively large because our DID design captures the dynamic and accumulating effects, rather than short-run effects attenuated by optimization frictions. Specifically, over seven years after the reform, workers gradually change their behavior in response to taxes while overcoming optimization frictions. These gradual responses lead to the accumulating effects, which are reflected in their relatively large elasticity. Therefore, our estimated elasticities will not be frictional ones \citep{Martinez2021} but can be interpreted as structural ones that are more relevant for long-run welfare \citep{Chetty2012, Chetty2011, Kleven2013, Kleven2023}.

For the medium-income group, the effects of taxes on wages are smaller and statistically insignificant. Their elasticity is 0.1, close to those estimated by \citet{Blomquist2010} for a similar sample (Swedish married males of working age in the 1980s). Larger tax effects among lower-income groups are also found by \citet{Zidar2019} using state-level data and variations in the United States. These findings on heterogeneous responses across income levels have implications for tax policies targeted toward low-income workers, such as the Earned Income Tax Credit in the United States. Given the relatively large elasticity and novel accumulating effects, the following analysis focuses on the low-income group and studies internal validity, the channels through which wages respond to taxes, and the responses of annual earnings and labor supply.

\subsection{Threats to identification}
This subsection shows that the estimation results for the low-income group survive the following threats to identification: compositional changes of employed workers, violations of the exclusion restriction, income reallocation between spouses, bunching, and robustness to alternative definitions of the low-income group.

\paragraph{Compositional changes.}
Recall that our sample consists of workers employed in the baseline pre-reform year 1986. Imposing a sample restriction that requires workers to be employed after the reform would introduce selection bias, as post-reform employment status is itself an outcome affected by the treatment---a bad control problem \citep{Angrist2009}. Consequently, outside of 1986, workers may be non-employed, which leads to missing wages, or may drop from the sample due to reasons such as death or emigration. To account for this attrition, we transform our unbalanced panel into a quasi-balanced panel by filling in each missing observation with an employment dummy set to zero. In this quasi-balanced panel, the filled-in observations have only the individual ID, year variable, and employment dummy, with all other variables missing.

\begin{figure}[!htbp]\centering
\caption{Employment dynamics}
\label{fig:emp_rate}
\medskip
\includegraphics[scale=0.6]{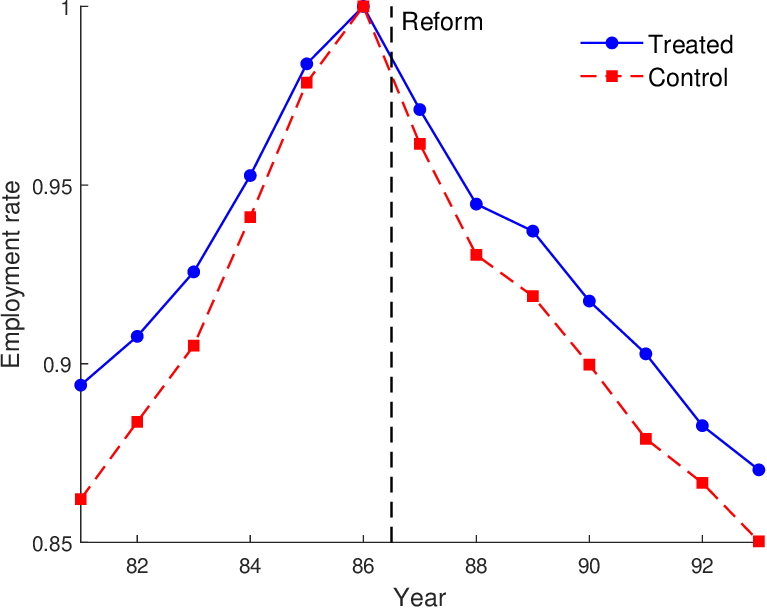}\hfill\includegraphics[scale=0.6]{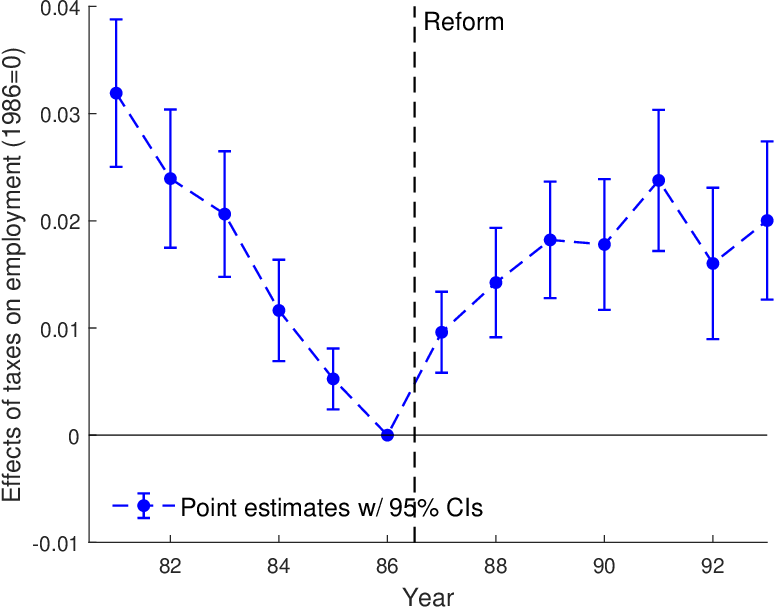}
\medskip
\caption*{\footnotesize Notes: The figure presents employment dynamics for the low-income group ($120,000 \leq \text{LI}_{i86} < 160,000$). Outcome $Y_{it}$ is a dummy variable indicating whether individual $i$ is employed ($Y_{it} = 1$) or not ($Y_{it} = 0$) in year $t$. The left panel plots $\overline{Y_{t}}$ for $t = 81, ..., 93$ by treatment status, where $\overline{Y_{t}}$ denotes mean $Y_{it}$ over $i$. The sample is males who, in 1986, were (i) younger than 50 years old and (ii) employed on the 28th of November. Furthermore, in 1986, (iii) they were married, and (iv) their wives had (strictly) positive labor income. The treated and control individuals are defined by the treatment assignment \eqref{eq:treated}. The right panel plots the point estimates of $\beta_t$ for $t=81, ..., 93$ with their 95\% confidence intervals from the two-way fixed effect (linear probability) model specified by Equation \eqref{eq:DID}. Standard errors are clustered at the individual level.}
\end{figure}

Figure \ref{fig:emp_rate} presents employment dynamics for the low-income group. Outcome $Y_{it}$ is the dummy variable indicating whether individual $i$ is employed ($Y_{it} = 1$) or not ($Y_{it} = 0$) in year $t$. The left panel shows high employment rates, higher than 85 percent, before and after the reform. The right panel plots DID coefficients $\widehat{\beta_t}$ from a linear probability model with the same specification as Equation \eqref{eq:DID} \citep{Lechner2010}. The statistically significant pre-reform effects ($\widehat{\beta_{81}}$, ..., $\widehat{\beta_{85}}$) suggest a violation of the parallel trends assumption, implying that the post-reform effects ($\widehat{\beta_{87}}$, ..., $\widehat{\beta_{93}}$) cannot be interpreted causally. Therefore, the post-reform employment rates between the treated and control individuals differ statistically, but not causally, only by 1--2 percentage points. Note that this small difference does not necessarily threaten identification: if exits from employment are random, the composition of employed workers remains constant over time, which validates our DID design. For identification, a key concern is not the quantity but the quality of employed workers: if the treated (control) individuals with higher (lower, respectively) wages exit employment, these compositional changes of employed workers will create the spurious negative effects of taxes on wages.

\begin{figure}[!htbp]\centering
\caption{Wage responses by the low-income group}
\label{fig:wage_low}
\medskip
\includegraphics[scale=0.6]{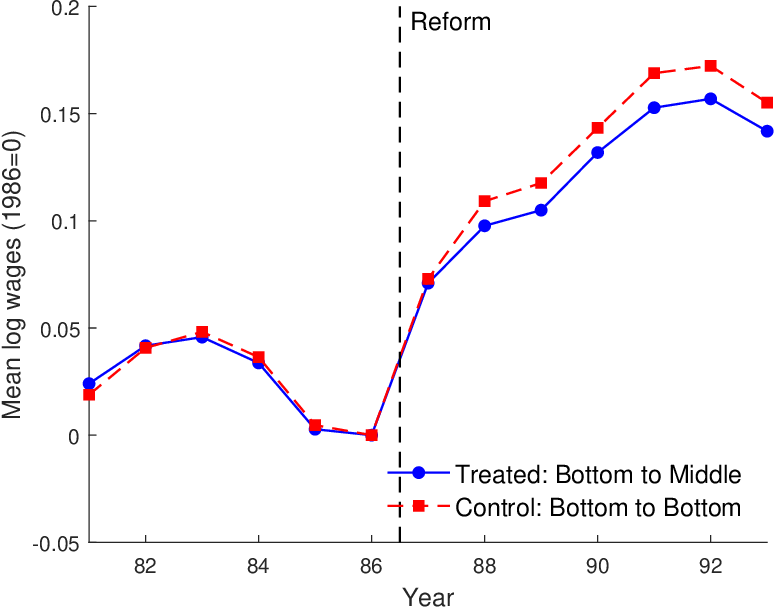}\hfill\includegraphics[scale=0.6]{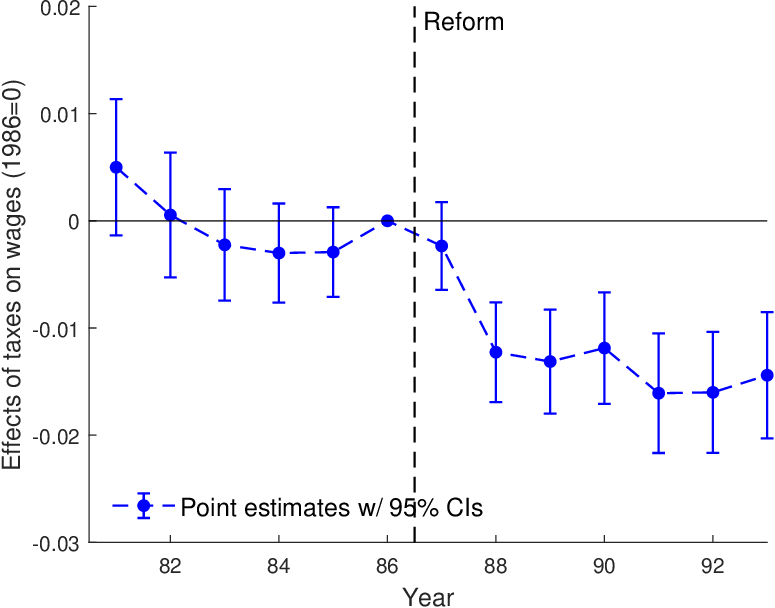}
\medskip
\caption*{\footnotesize Notes: The figure presents wage responses by the low-income group ($120,000 \leq \text{LI}_{i86} < 160,000$). Outcome $Y_{it}$ is the $\log$ of real gross hourly wages for a November job that individual $i$ holds in year $t$. The left panel plots $\overline{Y_{t}} - \overline{Y_{86}}$ for $t = 81, ..., 93$ by treatment status, where $\overline{Y_{t}}$ denotes mean $Y_{it}$ over $i$. The sample is males who, in 1986, were (i) younger than 50 years old and (ii) employed on the 28th of November. Furthermore, in 1986, (iii) they were married, and (iv) their wives had (strictly) positive labor income. The treated and control individuals are defined by the treatment assignment \eqref{eq:treated}. The right panel plots the point estimates of $\beta_t$ for $t=81, ..., 93$ with their 95\% confidence intervals from the two-way fixed effect model specified by Equation \eqref{eq:DID}. Standard errors are clustered at the individual level.}
\end{figure}

\begin{figure}[!htbp]\centering
\caption{Compositional changes of employed workers (measured by their 1986 wages)}
\label{fig:comp_change}
\medskip
\includegraphics[scale=0.8]{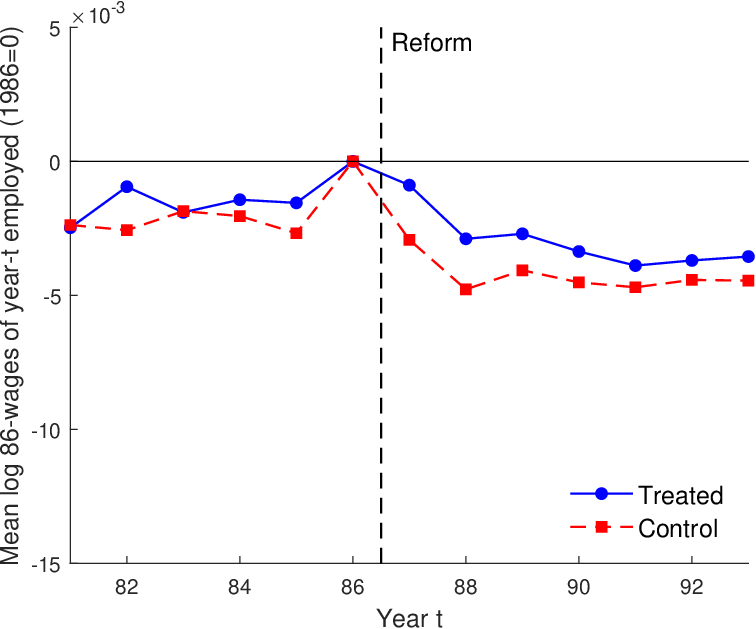}
\medskip
\caption*{\footnotesize Notes: The figure plots mean $\log$ wages in 1986 among workers employed in year $t$, by treatment status. The 1986 level is normalized to zero. The sample is males who, in 1986, were (i) younger than 50 years old and (ii) employed on the 28th of November. Furthermore, in 1986, (iii) they were married, and (iv) their wives had (strictly) positive labor income. The treated and control individuals are defined by the treatment assignment \eqref{eq:treated}. They are restricted to the low-income group ($120,000 \leq \text{LI}_{i86} < 160,000$).}
\end{figure}

To examine this concern, Figure \ref{fig:comp_change} plots mean $\log$ wages in 1986 among workers employed in year $t$, by treatment status. The 1986 level is normalized to zero. The treated and control individuals exhibit a similar pattern: compared to workers employed in 1986, workers employed outside of 1986 earned wages in 1986 that were lower by less than 0.5 percent. As displayed in Figure \ref{fig:wage_low}, the effect of taxes on wages is approximately $-0.015$, corresponding to the minimum value of the y-axis in Figure \ref{fig:comp_change}. If all of this tax effect were attributable to the compositional changes of employed workers, then Figure \ref{fig:comp_change} would show a post-reform difference of 0.015, with the treated individuals positioned below the control individuals. However, Figure \ref{fig:comp_change} actually shows a post-reform difference of only 0.001--0.002, with the treated individuals positioned \textit{above} the control individuals. Moreover, in contrast to the accumulating effects found in Figure \ref{fig:wage_low}, the 1986 wages evolve in parallel between the treated and control individuals. These results indicate that the compositional changes of employed workers (measured by their 1986 wages) occur to a slight but similar degree among the treated and control individuals, thereby not creating the spurious negative effects of taxes on wages.

Finally, Table \ref{tab:Covariate93} compares covariates in 1986 between workers employed in 1986 and workers employed in 1993, by treatment status. The comparison is based on normalized differences and reveals the same pattern as that of the 1986 wages: the compositional changes of employed workers (measured by their 1986 covariates) occur to a slight but similar degree among the treated and control individuals. To sum up, seven years after the reform, the employed treated individuals remain similar to the employed control individuals in terms of their 1986 wages and covariates (except for wives' labor income).

\paragraph{Exclusion restriction.}
Our empirical strategy leverages the variation in wives' labor income in 1986 ($\text{LI}_{i86}^\text{w}$) and effectively uses it as an instrumental variable in the DID design. We conduct a placebo test to verify the exclusion restriction, which requires that $\text{LI}_{i86}^\text{w}$ affects outcome dynamics only through the treatment (i.e., bracket locations and, thus, tax rates). Specifically, we show parallel pre- and post-reform outcome dynamics between placebo-treated and placebo-control individuals who differ exclusively in $\text{LI}_{i86}^\text{w}$ but face nearly identical tax rates (i.e., the absence of the treatment).

\begin{figure}[!htbp]\centering
\caption{Distributions of wives' pre-reform labor income $\text{LI}^\text{w}_{i86}$}
\label{fig:LI_density_wife}
\medskip
\includegraphics[scale=0.8]{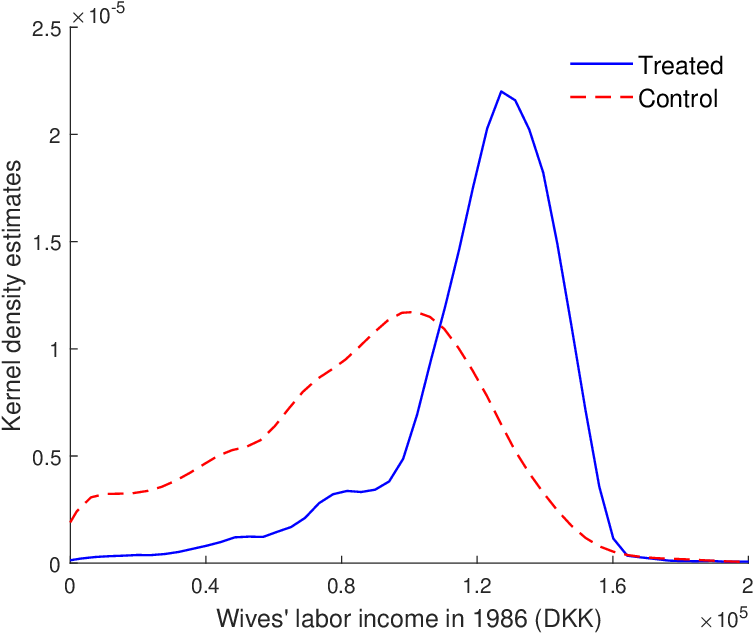}
\medskip
\caption*{\footnotesize Notes: The figure plots the kernel density estimates of wives' pre-reform labor income $\text{LI}^\text{w}_{i86}$ by treatment status. The estimation is based on a \texttt{ksdensity} function in MATLAB with default settings. DKK 1 in 1986 equals USD 0.3 in 2022. The sample is males who, in 1986, were (i) younger than 50 years old and (ii) employed on the 28th of November. Furthermore, in 1986, (iii) they were married, and (iv) their wives had (strictly) positive labor income. The treated and control individuals are defined by the treatment assignment \eqref{eq:treated}. They are restricted to the low-income group ($120,000 \leq \text{LI}_{i86} < 160,000$).}
\end{figure}

We construct placebo-treated and placebo-control individuals from the control individuals in the low-income group using their wives' labor income in 1986 ($\text{LI}_{i86}^\text{w}$). Figure \ref{fig:LI_density_wife} plots the kernel density estimates of $\text{LI}_{i86}^\text{w}$ by treatment status for the low-income group. As expected, the treated individuals have higher mean $\text{LI}_{i86}^\text{w}$ than the control individuals. If we select two control individuals with low $\text{LI}_{i86}^\text{w}$, they will be away from the middle-bracket cutoff and thus face similar tax rates over time. Based on this idea, we define a placebo group (composed of placebo-treated and placebo-control individuals) as follows:
\begin{align}\label{eq:placebo}
  \text{Placebo-treated: } & \ \text{Q}_{1} \leq \text{LI}_{i86}^\text{wC} < \text{Q}_{2} \notag \\
  \text{Placebo-control: } & \ \text{LI}_{i86}^\text{wC} < \text{Q}_{1},
\end{align}
where $\text{LI}_{i86}^\text{wC}$ denotes $\text{LI}_{i86}^\text{w}$ of the control individuals in the low-income group, and $\text{Q}_{1}$ and $\text{Q}_{2}$ denote the first and second quartiles of $\text{LI}_{i86}^\text{wC}$, respectively. ($\text{Q}_{1}$ $\approx$ DKK 60,000 and $\text{Q}_{2}$ $\approx$ DKK 90,000.)

Let us check their bracket locations and pre-reform covariates. First, Figure \ref{fig:bracket_placebo} and Figure \ref{fig:bracket_placebo_app} in Appendix \ref{app:fig} show that, unlike the low-income group, the placebo-treated and placebo-control individuals face nearly identical tax rates both before and after the reform; thus, their outcome dynamics represent those in the absence of the treatment. Next, Table \ref{tab:CovariatePlacebo} shows that, like the low-income group, the placebo-treated and placebo-control individuals are similar in pre-reform covariates except for wives' labor income $\text{LI}_{i86}^\text{w}$. The table also shows that the placebo group is similar to the low-income group. Furthermore, the placebo group has a larger normalized difference in $\text{LI}_{i86}^\text{w}$ (3.07) than the low-income group (1.12), thereby having a fair chance of exhibiting non-parallel outcome dynamics---evidence against the exclusion restriction. These observations suggest the validity of our placebo test.

\begin{figure}[!htbp]\centering
\caption{Wage responses by the placebo group}
\label{fig:wage_placebo}
\medskip
\includegraphics[scale=0.6]{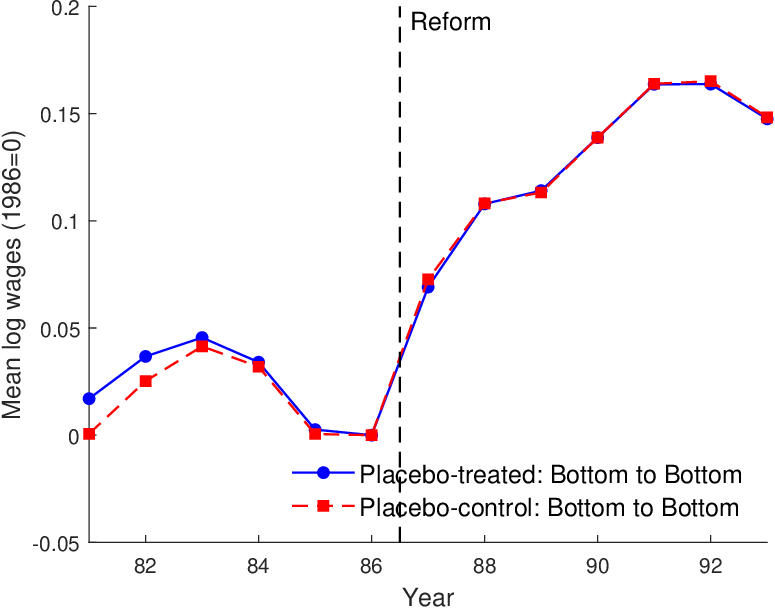}\hfill\includegraphics[scale=0.6]{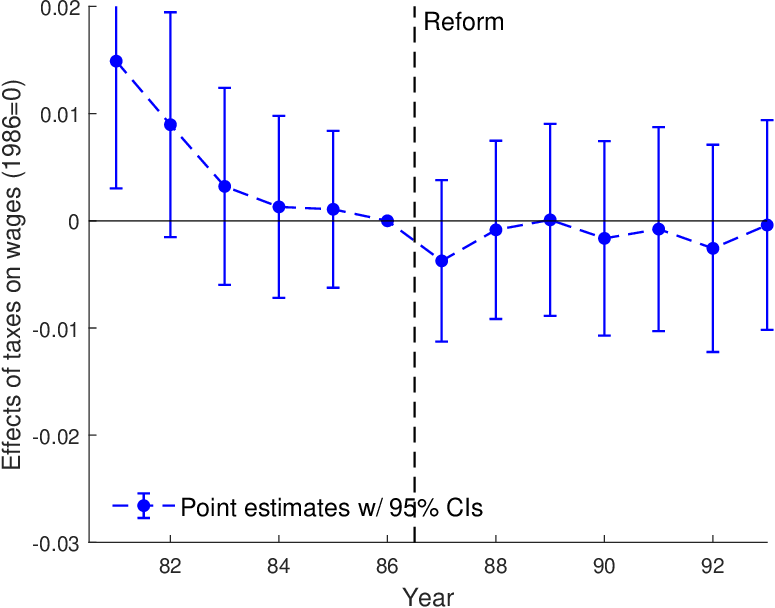}
\medskip
\caption*{\footnotesize Notes: The figure presents wage responses by the placebo group. Outcome $Y_{it}$ is the $\log$ of real gross hourly wages for a November job that individual $i$ holds in year $t$. The left panel plots $\overline{Y_{t}} - \overline{Y_{86}}$ for $t = 81, ..., 93$ by treatment status, where $\overline{Y_{t}}$ denotes mean $Y_{it}$ over $i$. The sample is males who, in 1986, were (i) younger than 50 years old and (ii) employed on the 28th of November. Furthermore, in 1986, (iii) they were married, and (iv) their wives had (strictly) positive labor income. They are restricted to the low-income group ($120,000 \leq \text{LI}_{i86} < 160,000$). The placebo-treated and placebo-control individuals are defined by the placebo assignment \eqref{eq:placebo}. The right panel plots the point estimates of $\beta_t$ for $t=81, ..., 93$ with their 95\% confidence intervals from the two-way fixed effect model specified by Equation \eqref{eq:DID}. Standard errors are clustered at the individual level.}
\end{figure}

Figure \ref{fig:wage_placebo} presents wage responses by the placebo-treated and placebo-control individuals. Except for $\widehat{\beta_{81}}$, the wage dynamics are parallel and not statistically different from each other before and after the reform; that is, $\text{LI}_{i86}^\text{w}$ does not affect the outcome dynamics in the absence of the treatment. Therefore, this placebo test provides evidence supporting the exclusion restriction.

\paragraph{Income reallocation.}
This placebo test also addresses another threat to identification. After the reform introduced the joint taxation, individual $i$'s tax rate depends not only on own income ($\text{LI}_{it}, \text{CI}_{it}, \text{D}_{it}$) but also on spousal income ($\text{LI}^\text{w}_{it}, \text{CI}^\text{w}_{it}, \text{D}^\text{w}_{it}$) for $t \geq 87$. Since the treated and control individuals in the low-income group differed (exclusively) in $\text{LI}^\text{w}_{i86}$ before the reform, they may have differential incentives to reallocate income (e.g., through labor supply) between spouses within a household to minimize tax liabilities after the reform. Such reallocation confounds our interpretation of wage responses as reflecting the effects of marginal tax rates, thereby threatening identification. 

We clarify two points regarding these differential incentives to reallocate income. First, such incentives may arise even without the joint taxation, provided that the treated and control individuals differ in $\text{LI}^\text{w}$ and optimize at the household level. However, this concern is ruled out by the parallel pre-reform trends shown in Figure \ref{fig:wage_low}. Thus, our primary concern is that the joint taxation induces these incentives. Second, such incentives persist even if the treated and control individuals face the same tax rate, as long as they differ in $\text{LI}^\text{w}$ under the joint taxation. 

Our placebo test finds no effect of income reallocation. Since the placebo-treated and placebo-control individuals face nearly identical tax rates and differ (exclusively) in $\text{LI}^\text{w}_{i86}$, their wage responses reflect the effects of income reallocation, rather than marginal tax rates. However, as shown in Figure \ref{fig:wage_placebo}, their wage dynamics remain parallel both before and after the reform, providing evidence that the introduction of the joint taxation does not affect the outcome dynamics through income reallocation.

\begin{figure}[!htbp]\centering
\caption{Density around the middle-bracket cutoff}
\label{fig:bunching}
\medskip
\includegraphics[scale=0.8]{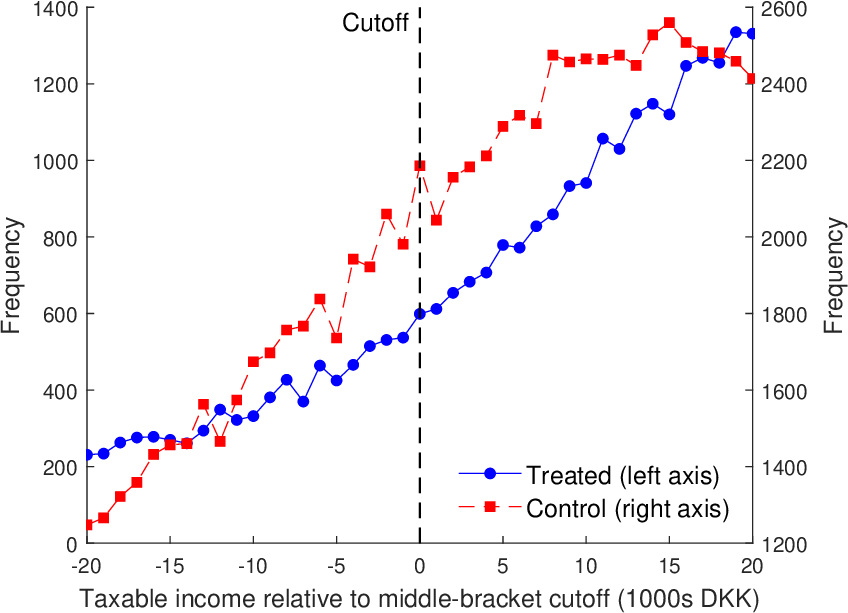}
\medskip
\caption*{\footnotesize Notes: The figure plots the frequencies of individuals by their taxable income relative to the middle-bracket cutoff in bins of DKK 1,000 for the post-reform period 1987--1993. For example, in 1987, the taxable income for the middle bracket was $\text{LI} + [\text{CI}>0]$, and the middle-bracket cutoff was DKK 130,000, as listed in Table \ref{tab:1987reform}. We deflate the taxable income and middle-bracket cutoffs to the 1986 price level. The figure consists of 41 bins in total: $[-20.5\text{k}, -19.5\text{k}), ..., [-1.5\text{k}, -500), [-500, 500), [500, 1.5\text{k}), ..., [19.5\text{k}, 20.5\text{k})$. DKK 1 in 1986 equals USD 0.3 in 2022. The sample is males who, in 1986, were (i) younger than 50 years old and (ii) employed on the 28th of November. Furthermore, in 1986, (iii) they were married, and (iv) their wives had (strictly) positive labor income. The treated and control individuals are defined by the treatment assignment \eqref{eq:treated}. They are restricted to the low-income group ($120,000 \leq \text{LI}_{i86} < 160,000$).}
\end{figure}

\paragraph{Bunching.}
As \citet{Kleven2014b} point out, quasi-experimental approaches exploiting tax reforms assume that individuals do not bunch at bracket cutoffs; otherwise, such bunching might bias estimates. Figure \ref{fig:bunching} plots the frequencies of individuals by their taxable income relative to the middle-bracket cutoff in bins of DKK 1,000 for the post-reform period 1987--1993. We deflate the taxable income and middle-bracket cutoffs to the 1986 price level. The figure shows no spikes around the cutoff; indeed, the lack of bunching among our sample (i.e., male wage-earners) at the middle-bracket cutoff aligns with \citet{Chetty2011} and \citet{LeMaire2013}, who find bunching primarily among females, self-employed workers, and at the top-bracket cutoff in Denmark.

\begin{figure}[!htbp]\centering
\caption{Wage responses by the four alternative low-income groups}
\label{fig:wage_robust}
\medskip
\caption*{(a) $115,000 \leq \text{LI}_{i86} < 160,000$}
{\includegraphics[scale=0.4]{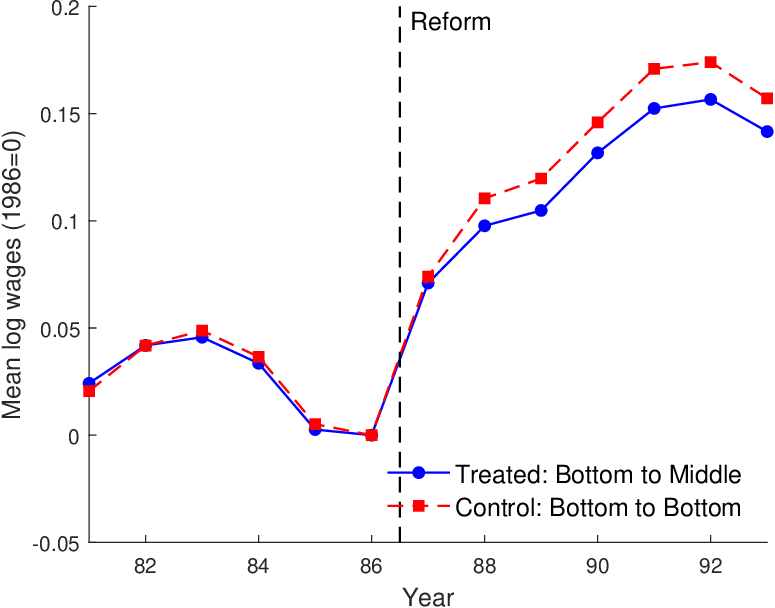}}\hspace{3em}
{\includegraphics[scale=0.4]{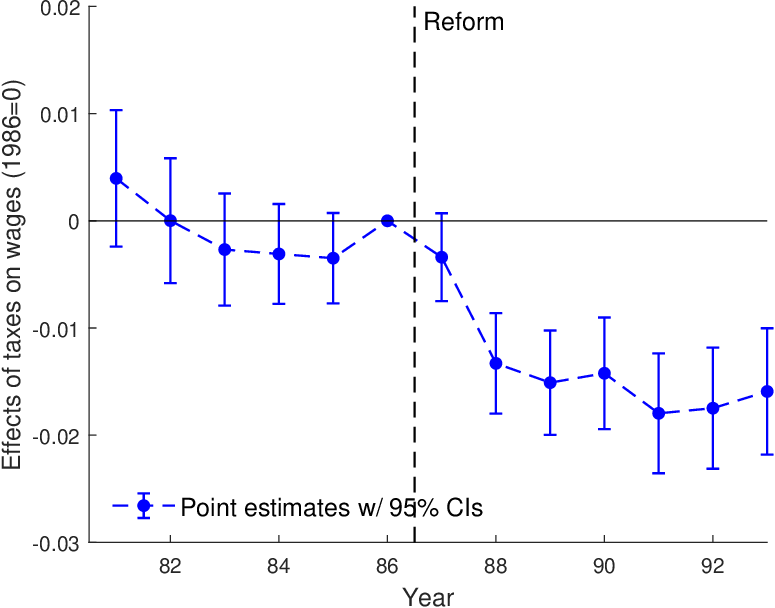}}

\bigskip

\caption*{(b) $125,000 \leq \text{LI}_{i86} < 160,000$}
{\includegraphics[scale=0.4]{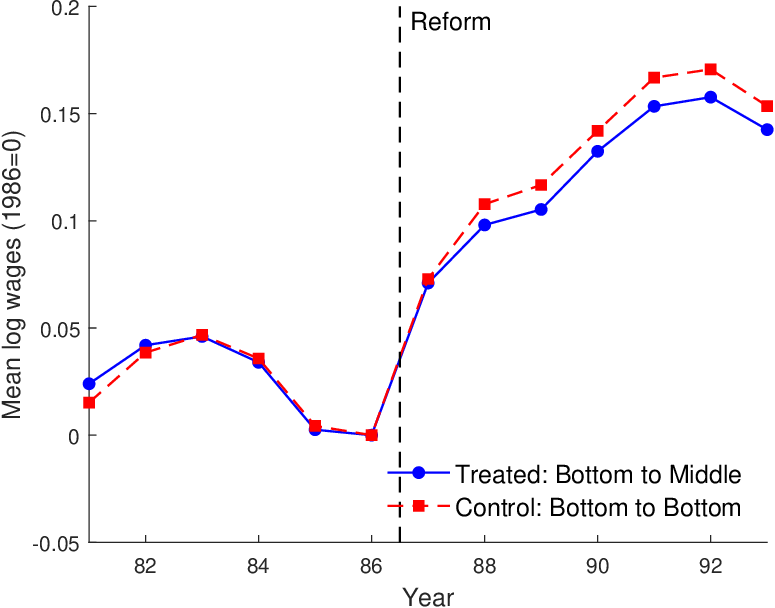}}\hspace{3em}
{\includegraphics[scale=0.4]{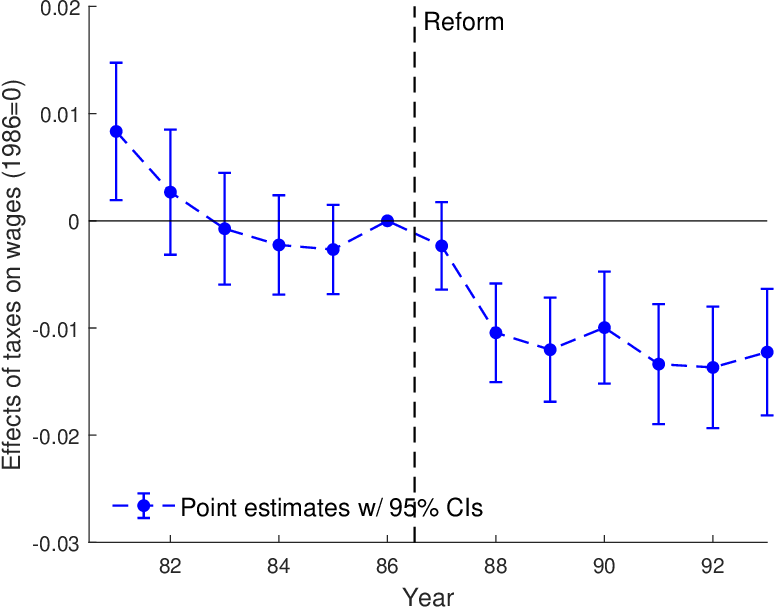}}

\bigskip

\caption*{(c) $120,000 \leq \text{LI}_{i86} < 155,000$}
{\includegraphics[scale=0.4]{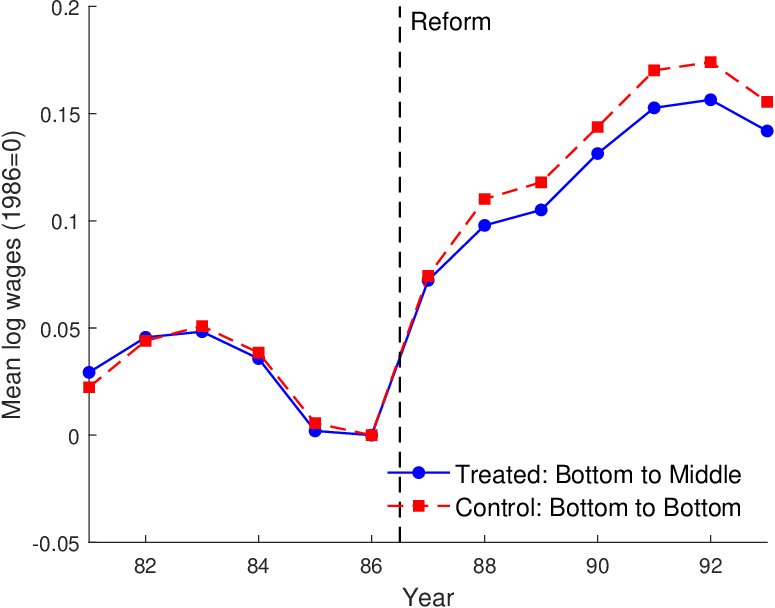}}\hspace{3em}
{\includegraphics[scale=0.4]{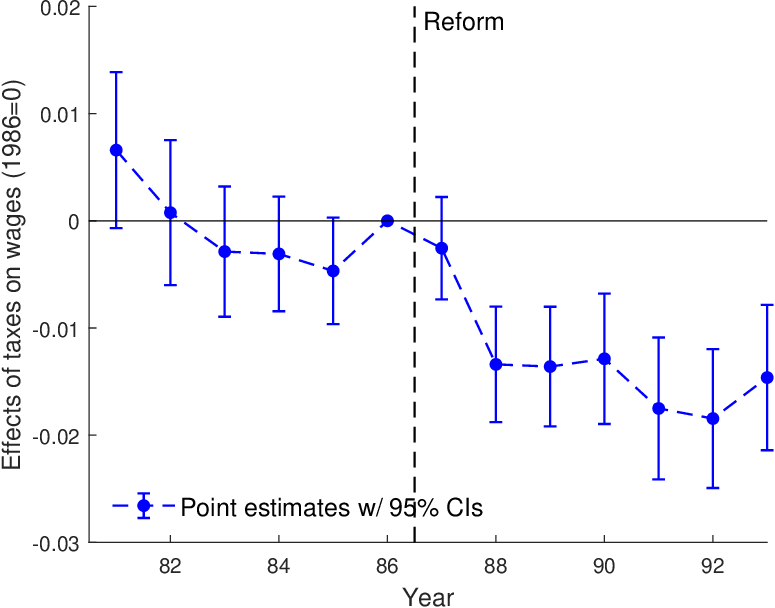}}

\bigskip

\caption*{(d) $120,000 \leq \text{LI}_{i86} < 165,000$}
{\includegraphics[scale=0.4]{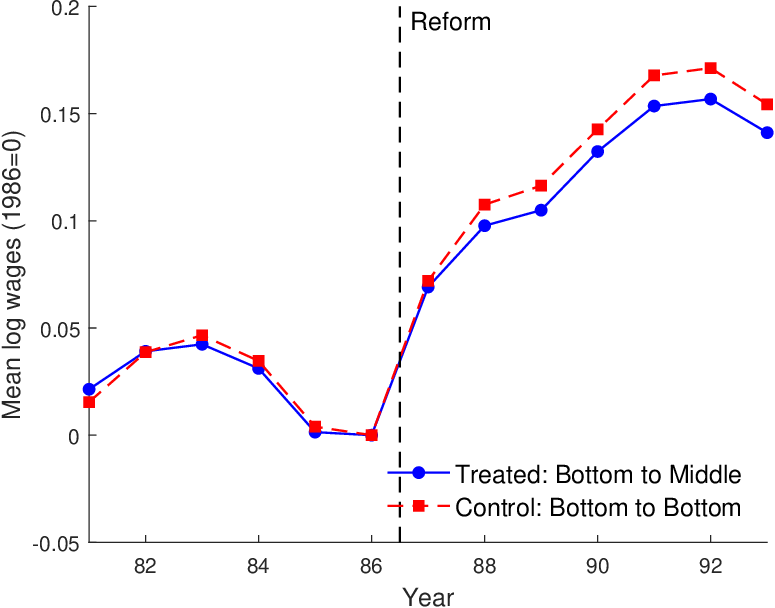}}\hspace{3em}
{\includegraphics[scale=0.4]{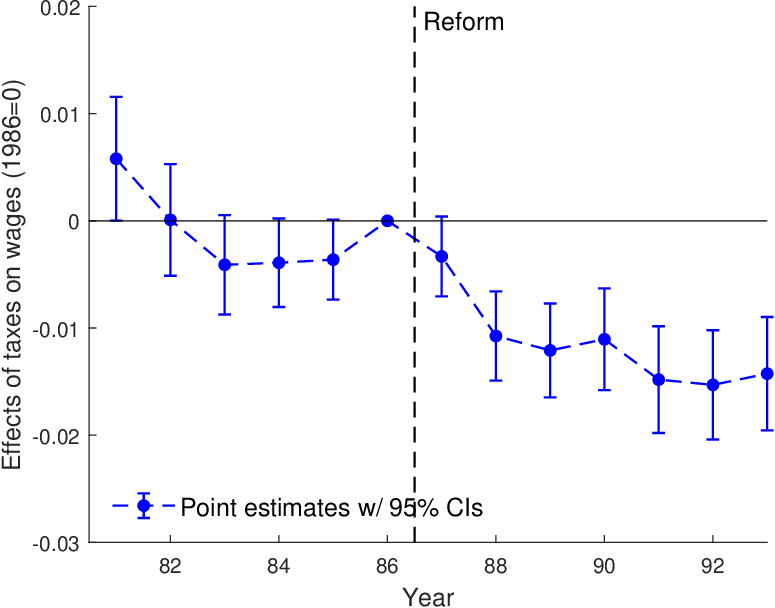}}
\medskip
\caption*{\footnotesize Notes: The figure presents wage responses by the four alternative low-income groups. Outcome $Y_{it}$ is the $\log$ of real gross hourly wages for a November job that individual $i$ holds in year $t$. The left panels plot $\overline{Y_{t}} - \overline{Y_{86}}$ for $t = 81, ..., 93$ by treatment status, where $\overline{Y_{t}}$ denotes mean $Y_{it}$ over $i$. The sample is males who, in 1986, were (i) younger than 50 years old and (ii) employed on the 28th of November. Furthermore, in 1986, (iii) they were married, and (iv) their wives had (strictly) positive labor income. The treated and control individuals are defined by the treatment assignment \eqref{eq:treated}. The right panels plot the point estimates of $\beta_t$ for $t=81, ..., 93$ with their 95\% confidence intervals from the two-way fixed effect model specified by Equation \eqref{eq:DID}. Standard errors are clustered at the individual level.}
\end{figure}

\paragraph{Robustness checks.}
Recall that the low-income group is defined as $120,000 \leq \text{LI}_{i86} < 160,000$. To check robustness to modest changes to this definition, we construct four alternative low-income groups by adding or subtracting 5,000 to either boundary of $120,000 \leq \text{LI}_{i86} < 160,000$. These four groups exhibit robust wage responses in Figure \ref{fig:wage_robust}, which presents graphical evidence and DID coefficients $\widehat{\beta_t}$. Their four implied elasticities, $\widehat{\epsilon}$ for each group, range from 0.327 ($\widehat{\text{s.e.}} = 0.058$) to 0.412 ($\widehat{\text{s.e.}} = 0.064$). These results align with Figure \ref{fig:wage_low} and $\widehat{\epsilon} = 0.369$ (0.065) of the original low-income group.

\subsection{Two channels: promotions and job-to-job transitions}
This subsection studies two channels through which wages can respond to taxes: promotions and job-to-job transitions. Workers respond to lower marginal tax rates potentially by accumulating human capital or working harder, both of which will lead to promotions, or by searching for higher-paying jobs, which will lead to job-to-job transitions. These dynamic channels can explain the negative and accumulating effects of taxes on wages; see \citet{Kleven2023} for similar channels.

\paragraph{Promotions.}
We examine promotions using information on occupational ranks. Statistics Denmark classifies employed workers into six ranked categories. In 1986, most workers in the low-income group fell into one of the bottom three categories: unskilled (34 percent), skilled (29 percent), and low-level white-collar (25 percent). We construct two outcome variables. First, we aggregate the six categories into two by creating a dummy variable that equals one if workers are ranked as skilled or higher; otherwise, it equals zero. Workers with this dummy variable equal to one are referred to as skilled hereafter in this paper. Second and similarly, we create another dummy variable that equals one if workers are ranked as low-level white-collar or higher; otherwise, it equals zero. Workers with this second dummy variable equal to one are referred to as white-collar.

\begin{figure}[!htbp]\centering
\caption{Promotions}
\label{fig:skilled}
\medskip
\caption*{(a) Being a skilled worker ($Y_{it} = 1$) or not ($Y_{it} = 0$)}
{\includegraphics[scale=0.6]{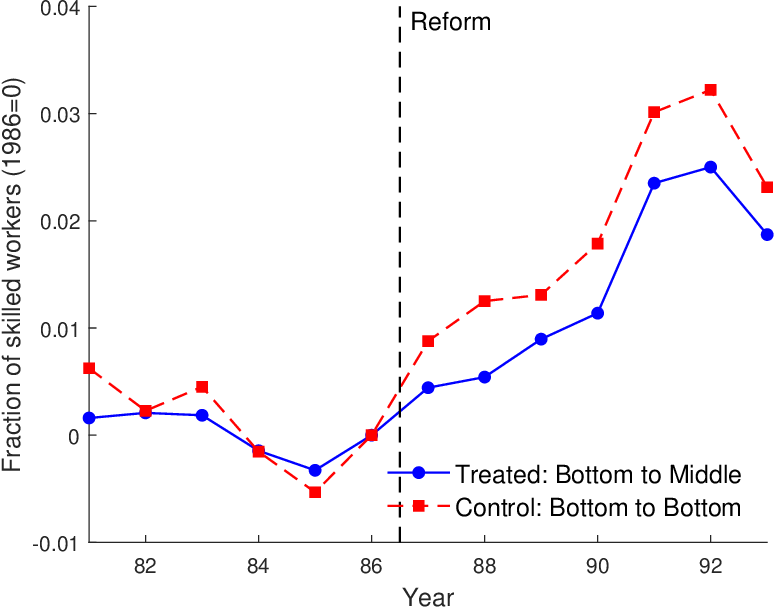}}\hfill
{\includegraphics[scale=0.6]{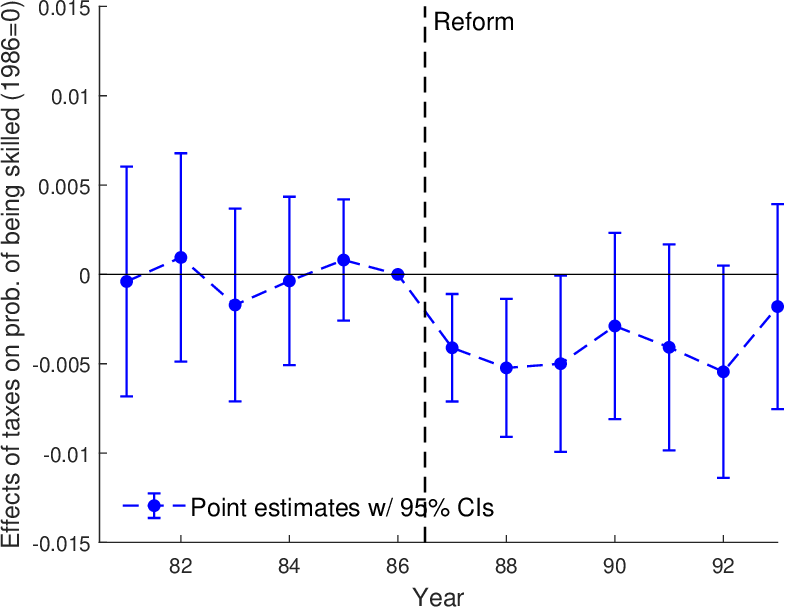}}

\bigskip

\caption*{(b) Being a white-collar worker ($Y_{it} = 1$) or not ($Y_{it} = 0$)}
{\includegraphics[scale=0.6]{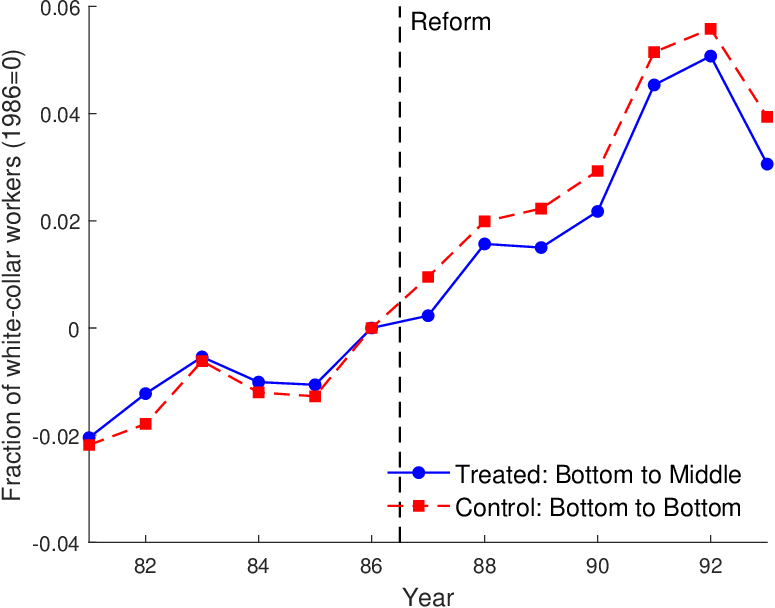}}\hfill
{\includegraphics[scale=0.6]{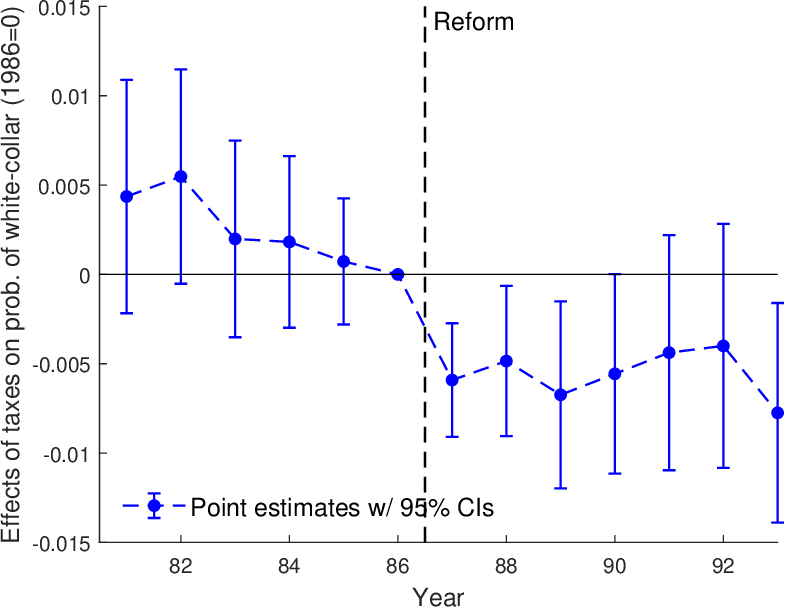}}
\medskip
\caption*{\footnotesize Notes: The figure presents promotions of the low-income group ($120,000 \leq \text{LI}_{i86} < 160,000$). Outcome $Y_{it}$ is a dummy variable indicating whether individual $i$ in year $t$ is skilled ($Y_{it} = 1$) or not ($Y_{it} = 0$) (in the top panels), or white-collar ($Y_{it} = 1$) or not ($Y_{it} = 0$) (in the bottom panels). The left panels plot $\overline{Y_{t}} - \overline{Y_{86}}$ for $t = 81, ..., 93$ by treatment status, where $\overline{Y_{t}}$ denotes mean $Y_{it}$ over $i$. The sample is males who, in 1986, were (i) younger than 50 years old and (ii) employed on the 28th of November. Furthermore, in 1986, (iii) they were married, and (iv) their wives had (strictly) positive labor income. The treated and control individuals are defined by the treatment assignment \eqref{eq:treated}. The right panels plot the point estimates of $\beta_t$ for $t=81, ..., 93$ with their 95\% confidence intervals from the two-way fixed effect (linear probability) model specified by Equation \eqref{eq:DID}. Standard errors are clustered at the individual level.}
\end{figure}

Like the wage responses, we provide graphical evidence, DID coefficients $\widehat{\beta_t}$, and implied semi-elasticities $\widehat{\epsilon}$. The left panels of Figure \ref{fig:skilled} plot the fractions of skilled workers (in the top panel) and white-collar workers (in the bottom panel) in year $t$ by treatment status. The two outcomes exhibit similar patterns. After the reform, the treated individuals are approximately 0.5 percentage points less likely to be promoted to skilled or white-collar workers than the control individuals; thus, higher marginal tax rates discourage promotions, arguably by reducing incentives to accumulate human capital or work harder. The right panels plot the DID coefficients $\widehat{\beta_t}$ from the same model as Equation \eqref{eq:DID}, where outcome $Y_{it}$ is the dummy variable indicating whether individual $i$ in year $t$ is skilled ($Y_{it} = 1$) or not ($Y_{it} = 0$) (in the top panel), or white-collar ($Y_{it} = 1$) or not ($Y_{it} = 0$) (in the bottom panel). Approximately half of the post-reform effects are statistically significant. For the first outcome (i.e., skilled or not), the implied semi-elasticity $\widehat{\epsilon}$ is 0.149 ($\widehat{\text{s.e.}} = 0.077$) with a $p$-value of 0.05. For the second outcome (i.e., white-collar or not), $\widehat{\epsilon}$ is 0.226 ($\widehat{\text{s.e.}} = 0.081$) and thus significant (refer to Table \ref{tab:elasticity} for a summary).

To the best of our knowledge, this paper provides the first quasi-experimental evidence on the negative effects of taxes on promotions. Our findings lend empirical support to recent research on optimal taxation with human capital accumulation or career effects \citep{Stantcheva2020}. In addition, our findings complement structural approaches that study the effects of income taxes on wages and hours by estimating learning-by-doing models \citep{Keane2012, Keane2015} or Ben-Porath models \citep{Heckman1998, Heckman1999}.

\paragraph{Job-to-job transitions.}
We define a job-to-job transition (JJT) between two consecutive years $t-1$ and $t$ based on three conditions: (i) workplace IDs differ between $t-1$ and $t$, (ii) wages are higher in $t$ than in $t-1$, and (iii) a worker is unemployed in neither $t-1$ nor $t$. In the first condition, we use workplace IDs rather than firm IDs due to data limitations. Our definition of JJTs is thus somewhat broad because it includes internal transfers involving workplace changes. Next, we impose the second condition to focus on workers climbing up job/wage ladders; in Denmark, 35 percent of job changes result in wage cuts \citep{Jolivet2006}. Finally, the third condition excludes involuntary job changes caused by layoffs. In this paper, unemployment is defined as receiving unemployment benefits.

Our outcome is the probability of making at least one JJT between 81 and year $t$ ($t = 82, ..., 93$). We clarify two points. First, recall that we observe only a November job for each worker and year; thus, we observe whether a worker makes zero or one JJT between two consecutive years $t-1$ and $t$. We calculate the number of JJTs between 81 and $t$ by summing up the number of JJTs between 81 and 82, between 82 and 83, ..., and between $t-1$ and $t$. Second, as predicted by a standard on-the-job search model, workers are less likely to make JJTs once they have settled into high-paying jobs. Therefore, to capture worker dynamics along their job/wage ladders, we examine not recent JJTs between $t-1$ and $t$, but rather whether workers have made any JJTs between $81$ and $t$, i.e., since the beginning of the sample period.

\begin{figure}[!htbp]\centering
\caption{Job-to-job transitions}
\label{fig:job-change}
\medskip
\includegraphics[scale=0.6]{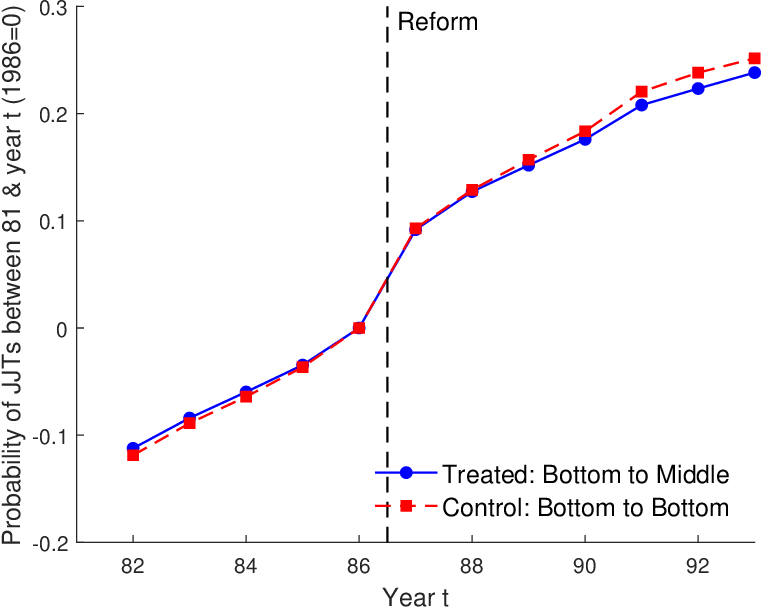}\hfill\includegraphics[scale=0.6]{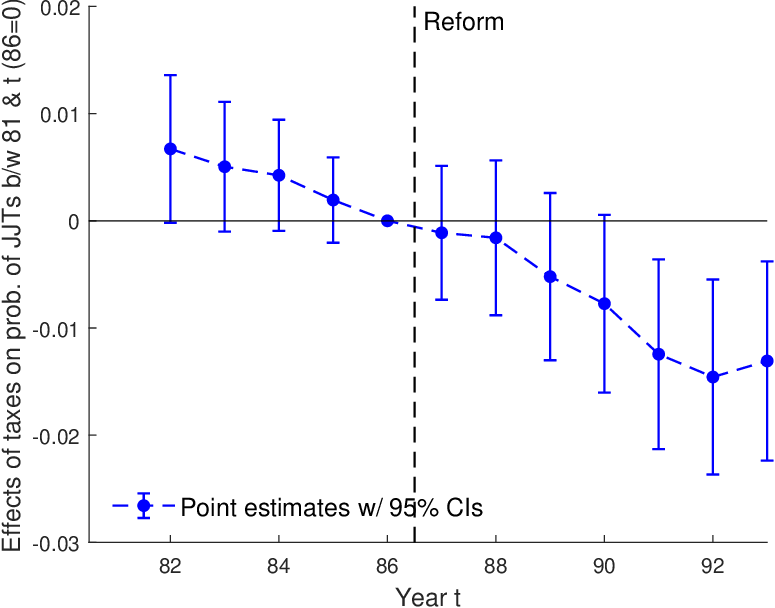}
\medskip
\caption*{\footnotesize Notes: The figure presents job-to-job transitions (JJTs) by the low-income group ($120,000 \leq \text{LI}_{i86} < 160,000$). Outcome $Y_{it}$ is a dummy variable indicating whether individual $i$ makes at least one JJT between 81 and year $t$ ($Y_{it} = 1$) or not ($Y_{it} = 0$). The left panel plots $\overline{Y_{t}} - \overline{Y_{86}}$ for $t = 82, ..., 93$ by treatment status, where $\overline{Y_{t}}$ denotes mean $Y_{it}$ over $i$. The sample is males who, in 1986, were (i) younger than 50 years old and (ii) employed on the 28th of November. Furthermore, in 1986, (iii) they were married, and (iv) their wives had (strictly) positive labor income. The treated and control individuals are defined by the treatment assignment \eqref{eq:treated}. The right panel plots the point estimates of $\beta_t$ for $t=82, ..., 93$ with their 95\% confidence intervals from the two-way fixed effect (linear probability) model specified by Equation \eqref{eq:DID}. Standard errors are clustered at the individual level.}
\end{figure}

We provide graphical evidence, DID coefficients $\widehat{\beta_t}$, and an implied semi-elasticity $\widehat{\epsilon}$. The left panel of Figure \ref{fig:job-change} plots the probability of making at least one JJT between 81 and year $t$ ($t = 82, ..., 93$) by treatment status. After the reform, the treated individuals are approximately one percentage point less likely to make JJTs than the control individuals; thus, higher marginal tax rates discourage JJTs, arguably by reducing incentives to search for higher-paying jobs. The right panel plots the DID coefficients $\widehat{\beta_t}$ from the same model as Equation \eqref{eq:DID}, where outcome $Y_{it}$ is a dummy variable indicating whether individual $i$ makes at least one JJT between 81 and year $t$ ($Y_{it} = 1$) or not ($Y_{it} = 0$). The last three post-reform effects ($\widehat{\beta_{91}}$, $\widehat{\beta_{92}}$, $\widehat{\beta_{93}}$) are statistically significant. The implied semi-elasticity $\widehat{\epsilon}$ is 0.370 ($\widehat{\text{s.e.}} = 0.144$) and thus significant.

Although \citet{Gentry2004} also find negative effects of taxes on JJTs, we provide the first quasi-experimental evidence. Our findings lend empirical support to welfare analyses based on job search models using structural approaches \citep{Kreiner2015, Shephard2017} or sufficient statistics approaches \citep{Chetty2008, Kroft2020}. Finally, JJTs are closely related to location choices, i.e., domestic or international migration. Given recent interest in the effects of taxes on migration \citep{Kleven2020}, exploring the interaction between JJTs and migration offers an interesting avenue for future research.

\subsection{Comparing responses of hourly wages, annual earnings, and labor supply}
\paragraph{Annual earnings.}
We use annual earnings from a November job as an outcome variable. These earnings correspond to the numerator in our definition of hourly wages. Note that labor income (LI) includes annual earnings from both November and non-November jobs. We use annual earnings from a November job, rather than LI, for two reasons. First, these earnings are, by definition, decomposed into hourly wages times annual hours worked for that job, which facilitates the comparison of their elasticities. Second and relatedly, since LI includes earnings from non-November jobs, it is inconsistent with our focus on November jobs.

\begin{figure}[!htbp]\centering
\caption{Annual earning responses}
\label{fig:earn}
\medskip
\includegraphics[scale=0.6]{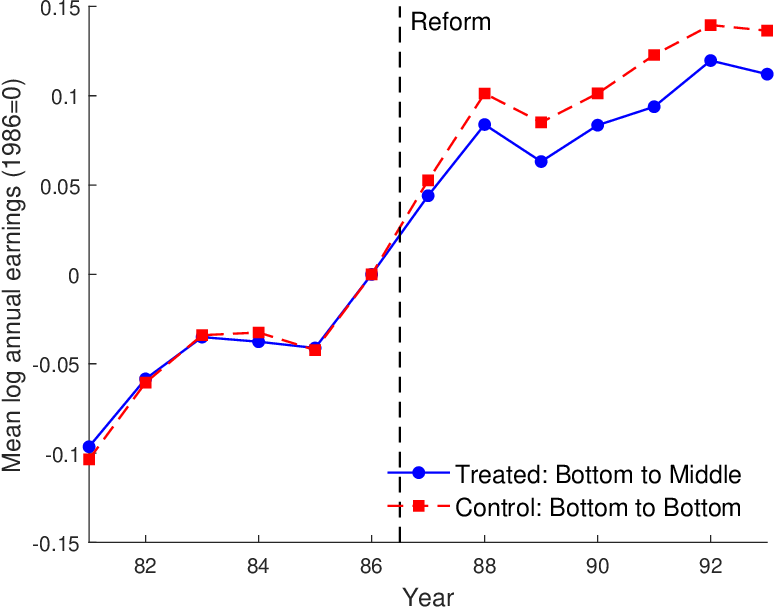}\hfill\includegraphics[scale=0.6]{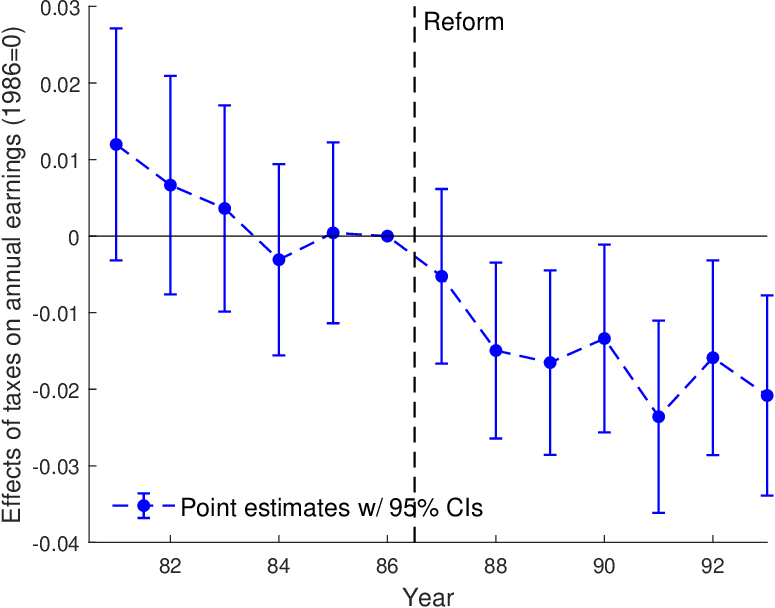}
\medskip
\caption*{\footnotesize Notes: The figure presents annual earning responses by the low-income group ($120,000 \leq \text{LI}_{i86} < 160,000$). Outcome $Y_{it}$ is the $\log$ of real gross annual earnings from a November job that individual $i$ holds in year $t$. The left panel plots $\overline{Y_{t}} - \overline{Y_{86}}$ for $t = 81, ..., 93$ by treatment status, where $\overline{Y_{t}}$ denotes mean $Y_{it}$ over $i$. The sample is males who, in 1986, were (i) younger than 50 years old and (ii) employed on the 28th of November. Furthermore, in 1986, (iii) they were married, and (iv) their wives had (strictly) positive labor income. The treated and control individuals are defined by the treatment assignment \eqref{eq:treated}. The right panel plots the point estimates of $\beta_t$ for $t=81, ..., 93$ with their 95\% confidence intervals from the two-way fixed effect model specified by Equation \eqref{eq:DID}. Standard errors are clustered at the individual level.}
\end{figure}

Figure \ref{fig:earn} presents annual earning responses. Outcome $Y_{it}$ is the $\log$ of real gross annual earnings from a November job that individual $i$ holds in year $t$. The left panel plots $\overline{Y_{t}} - \overline{Y_{86}}$ for $t = 81, ..., 93$ by treatment status, where $\overline{Y_{t}}$ denotes mean $Y_{it}$ over $i$. The right panel plots the DID coefficients $\widehat{\beta_t}$ from the same model as Equation \eqref{eq:DID}. Annual earnings exhibit responses similar to those of hourly wages displayed in Figure \ref{fig:wage_low}: insignificant pre-reform effects, followed by negative and accumulating post-reform effects. The implied elasticity $\widehat{\epsilon}$ is 0.519 ($\widehat{\text{s.e.}} = 0.143$) and thus somewhat larger than the wage elasticity of 0.369 (0.065).

Our elasticity of annual earnings is closely related to the literature on the elasticity of taxable income. Specifically, \citet{Kleven2014b} estimate the elasticity of labor income using a series of Danish tax reforms, including the 1987 reform. They report in Table 7 that their elasticity estimates based solely on the 1987 reform are 0.1, which is smaller than our elasticity estimate of 0.5.

Two possible reasons for this difference in the magnitude of estimates are empirical strategies and sample selection. \citet{Kleven2014b} follow an approach developed by \citet{Gruber2002}, which often compares individuals with varying pre-reform income patterns while exploiting reform-induced variation in tax rates across all tax brackets. Although this approach uses broad samples, thereby providing a comprehensive view of tax effects, it faces challenges in validating the parallel trends assumption \citep{Jakobsen2022}. By contrast, our empirical strategy compares individuals with similar pre-reform income patterns while focusing on reform-induced movements between the bottom and middle brackets, which makes the parallel trends assumption more plausible. To achieve this transparent identification, we impose more restrictions on our sample than \citet{Kleven2014b}. 

\begin{figure}[!htbp]\centering
\caption{Labor supply responses}
\label{fig:hour_FTPT}
\medskip
\caption*{(a) Daily hours worked}
{\includegraphics[scale=0.6]{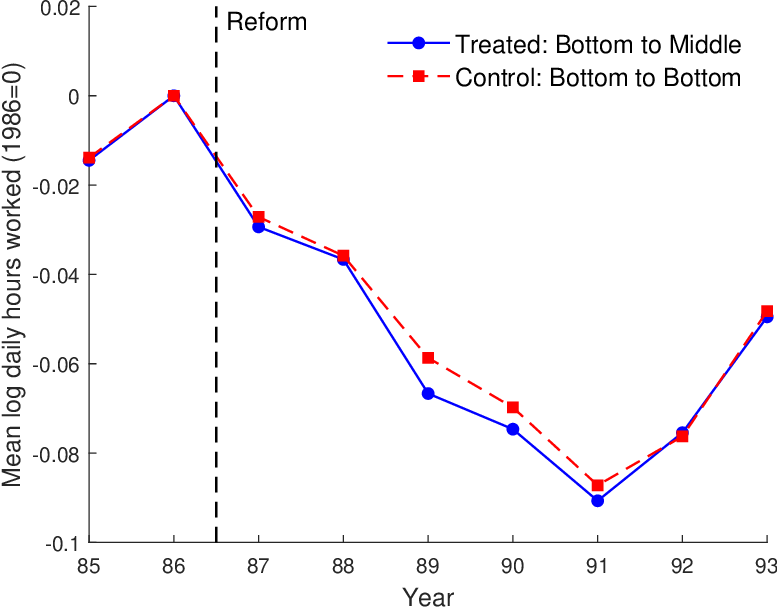}}\hfill
{\includegraphics[scale=0.6]{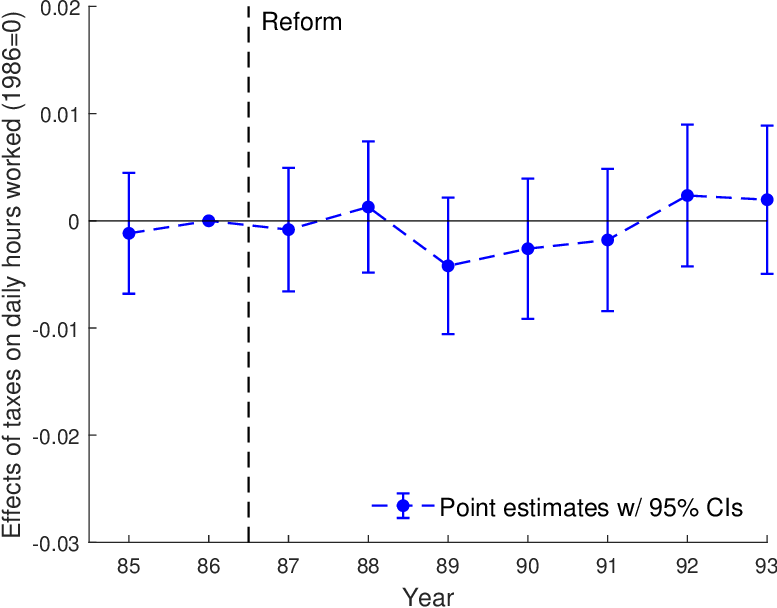}}

\bigskip

\caption*{(b) Annual hours worked}
{\includegraphics[scale=0.6]{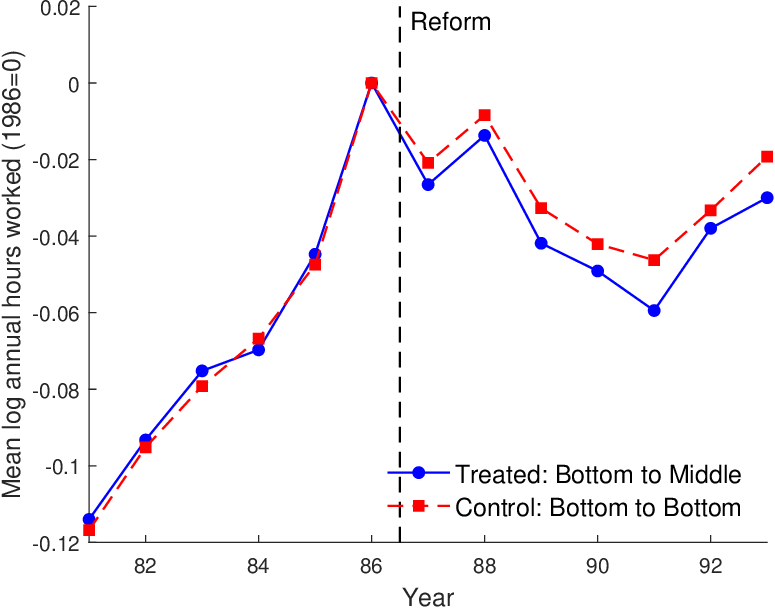}}\hfill
{\includegraphics[scale=0.6]{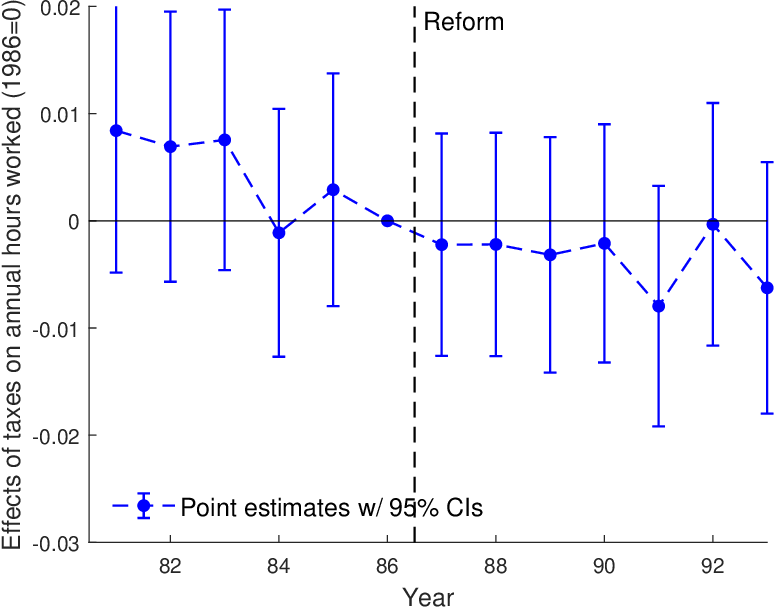}}
\medskip
\caption*{\footnotesize Notes: The figure presents labor supply responses by the low-income group ($120,000 \leq \text{LI}_{i86} < 160,000$). Outcome $Y_{it}$ is the $\log$ of daily (in the top panels) or annual (in the bottom panels) hours worked for a November job that individual $i$ holds in year $t$. Daily hours worked are missing from 1981 to 1984 due to data limitations. The left panels plot $\overline{Y_{t}} - \overline{Y_{86}}$ for $t = 85, ..., 93$ (in the top panel) or $t = 81, ..., 93$ (in the bottom panel) by treatment status, where $\overline{Y_{t}}$ denotes mean $Y_{it}$ over $i$. The sample is males who, in 1986, were (i) younger than 50 years old and (ii) employed on the 28th of November. Furthermore, in 1986, (iii) they were married, and (iv) their wives had (strictly) positive labor income. The treated and control individuals are defined by the treatment assignment \eqref{eq:treated}. The right panels plot the point estimates of $\beta_t$ for $t=85, ..., 93$ (in the top panel) or $t = 81, ..., 93$ (in the bottom panel) with their 95\% confidence intervals from the two-way fixed effect model specified by Equation \eqref{eq:DID}. Standard errors are clustered at the individual level.}
\end{figure}

\paragraph{Labor supply.}
Figure \ref{fig:hour_FTPT} presents labor supply responses along the intensive margin. Outcome $Y_{it}$ is the $\log$ of daily (in the top panels) or annual (in the bottom panels) hours worked for a November job that individual $i$ holds in year $t$. Daily hours worked are missing from 1981 to 1984 due to data limitations. Annual hours worked correspond to the denominator in our definition of hourly wages. The left panels provide graphical evidence, whereas the right panels plot DID coefficients $\widehat{\beta_t}$ using the same y-axis scale as the figures related to wage responses (e.g., Figure \ref{fig:wage_low}). Daily and annual hours worked exhibit similar responses: neither the pre- nor post-reform effects statistically differ from zero. Each implied elasticity $\widehat{\epsilon}$ is 0.000 ($\widehat{\text{s.e.}} = 0.078$) for daily hours and 0.132 ($\widehat{\text{s.e.}} = 0.128$) for annual hours. These insignificant results align with existing findings for married males \citep{Meghir2010}.

\paragraph{Summing up.}
We found that the elasticities of hourly wages, annual earnings, daily hours, and annual hours (with respect to net-of-tax rates) are 0.369, 0.519, 0.000, and 0.132, respectively, with the latter two being statistically insignificant (refer to Table \ref{tab:elasticity} for a summary). We highlight two points. First, the labor supply elasticities are insignificant and smaller than the wage elasticity. This finding has implications for optimal taxation because the literature derives tax formulae primarily by focusing on labor supply responses \citep{Piketty2013}. Second, the elasticity of annual earnings---a key parameter in the literature \citep{Saez2012}---is, by definition, decomposed into the elasticity of hourly wages plus that of annual hours worked. Our estimates indicate that the elasticity of annual earnings (0.519) is primarily driven by that of hourly wages (0.369), rather than by that of annual hours worked (0.132, insignificant). 

Related to the second point, \citet{Martinez2021} also provide credible quasi-experimental evidence on both wage and labor supply responses. They exploit a large and salient tax holiday induced by a reform in Switzerland, during which earned income was untaxed. Given that the tax holiday was a one-time tax cut, their empirical settings are well-suited for identifying intertemporal responses. Using biennial repeated cross-sectional data, they find the positive but insignificant elasticities of both hourly wages and monthly hours (with respect to net-of-tax rates). With some evidence on bonus shifting, they conclude that the wage responses are likely driven by tax avoidance.

In contrast to \citet{Martinez2021}, the current paper estimates the dynamic and accumulating effects of taxes, rather than the intertemporal effect of a one-time tax cut. In addition to the positive and significant elasticity of hourly wages, we found that wages respond to taxes through promotions or job-to-job transitions, rather than through tax avoidance. Therefore, our anatomy of behavioral responses, inspired by \citet{Slemrod1996}, provides new insights that complement those of \citet{Martinez2021}.

\section{Conclusion}\label{sec:conclusion}
This paper provided quasi-experimental evidence on the effects of income taxes on gross hourly wages by utilizing administrative data and a tax reform in Denmark. Our findings among working married males are as follows. First, taxes have heterogeneous effects on wages across income levels. Low-income workers respond to taxes negatively and dynamically; their elasticity of wages with respect to net-of-tax rates is 0.4. For medium-income workers, the effects are smaller and statistically insignificant. Second, wages respond to taxes through promotions or job-to-job transitions. Third, neither daily nor annual hours worked significantly respond to taxes; consequently, annual earnings respond to taxes primarily through hourly wages, rather than through labor supply.

Finally, we briefly discuss external validity. Although Denmark differs from the United States in many dimensions, the two countries share similar patterns in the elasticity of taxable income \citep{Gruber2002, Kleven2014b}. Moreover, we found heterogeneity similar to that found by \citet{Zidar2019} for the United States: larger tax effects among lower-income groups. Therefore, we believe that our findings are also relevant to other countries, such as the United States.

\clearpage

\bibliography{Bibliography}

\clearpage

\section*{Tables}

\begin{table}[!htbp]\centering
\caption{Income concepts in the Danish income tax system}
\label{tab:IncomeDef}
\begin{tabular}{lcl}
\hline
Income concept & Acronym & Main items included\tabularnewline
\hline
\hline
Labor income & LI & Salary, wages, bonuses, fringe benefits\tabularnewline
Capital income & CI & Interest income $-$ interest on debt\tabularnewline
Deductions & D & Commuting, union fees, UI contributions\tabularnewline
\hline
\end{tabular}
\medskip
\caption*{\footnotesize Notes: The table is based on \citet{Kleven2014b}. For ease of exposition, we omit two income concepts of minor importance for our sample: personal income and stock income. These omitted income concepts are explained later in Appendix \ref{app:simulator}. However, we emphasize that we fully consider all income concepts when simulating tax liabilities for the empirical analysis in Section \ref{sec:emp_str}.}
\end{table} 

\begin{table}[!htbp]\centering
\caption{The Danish income tax system before and after the 1987 tax reform}
\label{tab:1987reform}
\begin{tabular}{llrrrclrrr}
\hline
 & \multicolumn{4}{c}{1986} &  & \multicolumn{4}{c}{1987}\tabularnewline
\cline{2-5} \cline{7-10}
Tax type & Base & Cutoff & Joint & Rate &  & Base & Cutoff & Joint & Rate\tabularnewline
\hline
\hline
Regional taxes & $\text{LI} + \text{CI} - \text{D}$ & 20,700 & No & 28.0 &  & $\text{LI} + \text{CI} - \text{D}$ & 21,200 & No & 29.0\tabularnewline
 &  &  &  &  &  &  &  &  & \tabularnewline
National taxes &  &  &  &  &  &  &  &  & \tabularnewline
Bottom bracket & $\text{LI} + \text{CI} - \text{D}$ & 23,200 & No & 19.9 &  & $\text{LI} + \text{CI} - \text{D}$ & 27,100 & No & 22.0\tabularnewline
Middle bracket & $\text{LI} + \text{CI} - \text{D}$ & 113,400 & No & 14.4 &  & $\text{LI} + [\text{CI}>0]$ & 130,000 & Yes & 6.0\tabularnewline
Top bracket & $\text{LI} + \text{CI} - \text{D}$ & 186,100 & No & 10.8 &  & $\text{LI} + [\text{CI}>60\text{k}]$ & 200,000 & No & 12.0\tabularnewline
\hline
\end{tabular}
\medskip
\caption*{\footnotesize Notes: All monetary values are in Danish Krone (DKK), and DKK 1 in 1986 equals USD 0.3 in 2022. The regional tax rate is the sum of the municipal, county, and Church tax rates. The Church taxes are minuscule and are paid only by members of the Church of Denmark (``Folkekirken''). The regional tax rates in the table are averages across municipalities. The bottom tax rate in 1986 includes social security contributions levied at a tax rate of 5.5 percent. ``Yes" in the ``Joint" column means that if married individuals are not liable for the middle taxes and have unused allowances (calculated as the bracket cutoff DKK 130,000 minus their taxable income $\text{LI} + [\text{CI}>0]$), these unused allowances can be transferred to their spouses.}
\end{table} 

\begin{table}[!htbp]\centering
\caption{Summary statistics of pre-reform covariates for each income group}
\label{tab:Covariate}
\begin{tabular}{lrrrcrrr}
\hline 
 & \multicolumn{3}{c}{Low-income} &  & \multicolumn{3}{c}{Medium-income}\tabularnewline
\cline{2-4} \cline{6-8} 
Variable & Treated & Control & N. d. &  & Treated & Control & N. d.\tabularnewline
\hline 
\hline 
Labor income & 148,596 & 142,386 & 0.65 &  & 185,462 & 181,446 & 0.19\tabularnewline
 & (8,202) & (10,625) &  &  & (21,667) & (20,444) & \tabularnewline
 &  &  &  &  &  &  & \tabularnewline
Age & 36.8 & 36.1 & 0.11 &  & 37.0 & 36.5 & 0.08\tabularnewline
 & (6.7) & (7.2) &  &  & (6.1) & (6.5) & \tabularnewline
Number of children & 1.3 & 1.5 & $-$0.11 &  & 1.4 & 1.6 & $-$0.17\tabularnewline
 & (0.9) & (0.9) &  &  & (0.9) & (0.9) & \tabularnewline
Low education (\%) & 35.4 & 38.6 & $-$0.07 &  & 24.1 & 24.7 & $-$0.01\tabularnewline
 & (47.8) & (48.7) &  &  & (42.8) & (43.1) & \tabularnewline
Middle education (\%) & 57.7 & 56.4 & 0.03 &  & 62.7 & 64.0 & $-$0.03\tabularnewline
 & (49.4) & (49.6) &  &  & (48.4) & (48.0) & \tabularnewline
High education (\%) & 6.9 & 5.0 & 0.08 &  & 13.2 & 11.3 & 0.06\tabularnewline
 & (25.4) & (21.8) &  &  & (33.9) & (31.7) & \tabularnewline
Full-time job (\%) & 46.7 & 46.3 & 0.01 &  & 62.3 & 62.0 & 0.00\tabularnewline
 & (49.9) & (49.9) &  &  & (48.5) & (48.5) & \tabularnewline
Private-sector job (\%) & 70.5 & 68.5 & 0.04 &  & 67.8 & 68.5 & $-$0.01\tabularnewline
 & (45.6) & (46.4) &  &  & (46.7) & (46.5) & \tabularnewline
 &  &  &  &  &  &  & \tabularnewline
Capital income & $-$38,742 & $-$43,596 & 0.25 &  & $-$68,323 & $-$74,287 & 0.23\tabularnewline
 & (15,513) & (22,617) &  &  & (24,565) & (27,802) & \tabularnewline
Deductions & 10,647 & 10,354 & 0.04 &  & 14,908 & 14,724 & 0.01\tabularnewline
 & (6,605) & (7,903) &  &  & (11,730) & (14,322) & \tabularnewline
Capital income (wife) & $-$5,285 & $-$6,880 & 0.12 &  & $-$3,959 & $-$5,648 & 0.12\tabularnewline
 & (12,244) & (14,651) &  &  & (12,605) & (15,984) & \tabularnewline
Deductions (wife) & 7,623 & 8,108 & $-$0.06 &  & 8,530 & 8,309 & 0.02\tabularnewline
 & (6,344) & (9,454) &  &  & (8,275) & (10,003) & \tabularnewline
 &  &  &  &  &  &  & \tabularnewline
Labor income (wife) & 119,978 & 83,925 & 1.12 &  & 127,327 & 82,986 & 1.23\tabularnewline
 & (26,352) & (37,209) &  &  & (32,725) & (38,964) & \tabularnewline
\hline 
\hline 
Number of individuals & 11,642 & 28,338 &  &  & 21,211 & 16,303 & \tabularnewline
N. of observations (1981--1993) & 150,309 & 365,257 &  &  & 273,674 & 210,060 & \tabularnewline
\hline 
\end{tabular} 
\medskip
\caption*{\footnotesize Notes: The table lists the means of covariates in 1986 by treatment status for each income group. Standard deviations are in parentheses. ``N. d." denotes a normalized difference, defined as $\frac{\overline{X}_T - \overline{X}_C}{\sqrt{(S_T^2 + S_C^2) / 2}}$ for each covariate, where $\overline{X}_T$ and $\overline{X}_C$ are the means among the treated and control individuals respectively, and $S_T$ and $S_C$ are the corresponding standard deviations. As a general guideline, a covariate is considered as balanced between the treated and control individuals if its normalized difference is less than 0.25. All monetary values are in Danish Krone (DKK), and DKK 1 in 1986 equals USD 0.3 in 2022. Children are defined as individuals younger than 17 years old. Low education is defined as completing primary education. Middle education is defined as completing high school or vocational education. High education is defined as holding a bachelor's, master's, or Ph.D. degree. Full-time jobs are defined as working more than 30 hours per week. The sample is males who, in 1986, were (i) younger than 50 years old and (ii) employed on the 28th of November. Furthermore, in 1986, (iii) they were married, and (iv) their wives had (strictly) positive labor income. The treated and control individuals are defined by the treatment assignment \eqref{eq:treated}. They are restricted to the low-income ($120,000 \leq \text{LI}_{i86} < 160,000$) and medium-income ($160,000 \leq \text{LI}_{i86} < 280,000$) groups.}
\end{table} 

\begin{table}[!htbp]\centering
\caption{Elasticity of outcome $Y$ with respect to net-of-tax rates, and related results}
\label{tab:elasticity}
\begin{tabular}{llrrrrrr}
\hline 
Variable &  &  & Low-income &  &  & Medium-income & \tabularnewline
\hline 
\hline 
Continuous outcome $Y$ &  &  &  &  &  &  & \tabularnewline
\cline{1-1} 
Gross hourly wages &  &  &  &  &  &  & \tabularnewline
\qquad{}First stage &  &  &  &  &  &  & \tabularnewline
\qquad{}\qquad{}Coefficient on IV &  &  & 0.229 (0.003) &  &  & 0.158 (0.003) & \tabularnewline
\qquad{}\qquad{}$F$-statistic $>$ 104.7 &  & \multicolumn{3}{c}{Yes} & \multicolumn{3}{c}{Yes}\tabularnewline
\qquad{}TOT effect $\widehat{\beta^{\text{TOT}}}$ &  &  & $-$0.044 (0.008) &  &  & $-$0.015 (0.012) & \tabularnewline
\qquad{}Elasticity $\widehat{\epsilon}$ &  &  & 0.369 (0.065) &  &  & 0.112 (0.089) & \tabularnewline
Gross annual earnings &  &  &  &  &  &  & \tabularnewline
\qquad{}TOT effect $\widehat{\beta^{\text{TOT}}}$ &  &  & $-$0.062 (0.017) &  &  &  & \tabularnewline
\qquad{}Elasticity $\widehat{\epsilon}$ &  &  & 0.519 (0.143) &  &  &  & \tabularnewline
Daily hours worked &  &  &  &  &  &  & \tabularnewline
\qquad{}TOT effect $\widehat{\beta^{\text{TOT}}}$ &  &  & $-$0.000 (0.009) &  &  &  & \tabularnewline
\qquad{}Elasticity $\widehat{\epsilon}$ &  &  & 0.000 (0.078) &  &  &  & \tabularnewline
Annual hours worked &  &  &  &  &  &  & \tabularnewline
\qquad{}TOT effect $\widehat{\beta^{\text{TOT}}}$ &  &  & $-$0.016 (0.015) &  &  &  & \tabularnewline
\qquad{}Elasticity $\widehat{\epsilon}$ &  &  & 0.132 (0.128) &  &  &  & \tabularnewline
 &  &  &  &  &  &  & \tabularnewline
Binary outcome $Y$ &  &  &  &  &  &  & \tabularnewline
\cline{1-1} 
Being a skilled worker or not &  &  &  &  &  &  & \tabularnewline
\qquad{}TOT effect $\widehat{\beta^{\text{TOT}}}$ &  &  & $-$0.018 (0.009) &  &  &  & \tabularnewline
\qquad{}Semi-elasticity $\widehat{\epsilon}$ &  &  & 0.149 (0.077) &  &  &  & \tabularnewline
Being a white-collar worker or not &  &  &  &  &  &  & \tabularnewline
\qquad{}TOT effect $\widehat{\beta^{\text{TOT}}}$ &  &  & $-$0.027 (0.010) &  &  &  & \tabularnewline
\qquad{}Semi-elasticity $\widehat{\epsilon}$ &  &  & 0.226 (0.081) &  &  &  & \tabularnewline
Making job-to-job transitions or not &  &  &  &  &  &  & \tabularnewline
\qquad{}TOT effect $\widehat{\beta^{\text{TOT}}}$ &  &  & $-$0.044 (0.017) &  &  &  & \tabularnewline
\qquad{}Semi-elasticity $\widehat{\epsilon}$ &  &  & 0.370 (0.144) &  &  &  & \tabularnewline
\hline 
\hline 
Mechanical changes in net-of-tax rates &  &  &  &  &  &  & \tabularnewline
\qquad{}$\text{avg}[\ \Delta\log(1-\tau_{i}^{\text{Mech}})\ |\ \text{Treated}\ ]$ &  & \multicolumn{3}{c}{$-$0.164} & \multicolumn{3}{c}{$-$0.175}\tabularnewline
\qquad{}$\text{avg}[\ \Delta\log(1-\tau_{i}^{\text{Mech}})\ |\ \text{Control}\ ]$ &  & \multicolumn{3}{c}{$-$0.044} & \multicolumn{3}{c}{$-$0.043}\tabularnewline
\hline 
\end{tabular}
\medskip
\caption*{\footnotesize Notes: The table lists the main elasticity estimates, along with related results. Standard errors are in parentheses. We compute an elasticity in two steps. In the first step, we compute a TOT effect $\widehat{\beta^{\text{TOT}}}$ by estimating Equation \eqref{eq:TOT} while instrumenting $Post_t \cdot M_{it}$ with $Post_t \cdot Treated_i$. Note that \citet{Lee2022} show that a first-stage $F$-statistic greater than 104.7 ensures a correct size of 5\% for a two-sided $t$-test for a 2SLS coefficient in broad settings of just-identified, single IV cases. Since the first-stage results are almost identical across outcomes, we report them only for gross hourly wages. Standard errors are clustered at the individual level. In the second step, using Equation \eqref{eq:elasticity}, we compute $\widehat{\epsilon}$, the elasticity of outcome $Y$ with respect to net-of-tax rates. Although not reported in the table, the standard deviations of $\Delta \log(1-\tau_{i}^\text{Mech})$ are only approximately 10$^{-4}$; thus, we treat the mechanical changes in net-of-tax rates as constants. Standard errors are computed using the delta method. The sample is males who, in 1986, were (i) younger than 50 years old and (ii) employed on the 28th of November. Furthermore, in 1986, (iii) they were married, and (iv) their wives had (strictly) positive labor income. The treated and control individuals are defined by the treatment assignment \eqref{eq:treated}. They are restricted to the low-income ($120,000 \leq \text{LI}_{i86} < 160,000$) and medium-income ($160,000 \leq \text{LI}_{i86} < 280,000$) groups.}
\end{table} 

\begin{table}[!htbp]\centering
\caption{Summary statistics of pre-reform covariates among workers employed in 1986 or 1993}
\label{tab:Covariate93}
\begin{tabular}{lrrrcrrr}
\hline 
 & \multicolumn{3}{c}{Treated} &  & \multicolumn{3}{c}{Control}\tabularnewline
\cline{2-4} \cline{6-8} 
Variable & Empl. in 86 & Empl. in 93 & N. d. &  & Empl. in 86 & Empl. in 93 & N. d.\tabularnewline
\hline 
\hline 
Age & 36.8 & 36.6 & 0.04 &  & 36.1 & 35.9 & 0.03\tabularnewline
 & (6.7) & (6.6) &  &  & (7.2) & (7.1) & \tabularnewline
Number of children & 1.3 & 1.4 & $-$0.02 &  & 1.5 & 1.5 & $-$0.01\tabularnewline
 & (0.9) & (0.9) &  &  & (0.9) & (0.9) & \tabularnewline
Low education (\%) & 35.4 & 34.8 & 0.01 &  & 38.6 & 38.0 & 0.01\tabularnewline
 & (47.8) & (47.6) &  &  & (48.7) & (48.5) & \tabularnewline
Middle education (\%) & 57.7 & 57.8 & $-$0.00 &  & 56.4 & 56.7 & $-$0.01\tabularnewline
 & (49.4) & (49.4) &  &  & (49.6) & (49.5) & \tabularnewline
High education (\%) & 6.9 & 7.4 & $-$0.02 &  & 5.0 & 5.3 & $-$0.01\tabularnewline
 & (25.4) & (26.2) &  &  & (21.8) & (22.4) & \tabularnewline
Full-time job (\%) & 46.7 & 48.2 & $-$0.03 &  & 46.3 & 47.9 & $-$0.03\tabularnewline
 & (49.9) & (50.0) &  &  & (49.9) & (50.0) & \tabularnewline
Private-sector job (\%) & 70.5 & 68.6 & 0.04 &  & 68.5 & 66.3 & 0.05\tabularnewline
 & (45.6) & (46.4) &  &  & (46.4) & (47.3) & \tabularnewline
 &  &  &  &  &  &  & \tabularnewline
Labor income & 148,596 & 148,919 & $-$0.04 &  & 142,386 & 142,828 & $-$0.04\tabularnewline
 & (8,202) & (7,977) &  &  & (10,625) & (10,475) & \tabularnewline
Capital income & $-$38,742 & $-$38,775 & 0.00 &  & $-$43,596 & $-$43,800 & 0.01\tabularnewline
 & (15,513) & (14,996) &  &  & (22,617) & (22,014) & \tabularnewline
Deductions & 10,647 & 10,496 & 0.02 &  & 10,354 & 10,094 & 0.03\tabularnewline
 & (6,605) & (6,454) &  &  & (7,903) & (7,457) & \tabularnewline
Capital income (wife) & $-$5,285 & $-$5,259 & $-$0.00 &  & $-$6,880 & $-$6,827 & $-$0.00\tabularnewline
 & (12,244) & (11,621) &  &  & (14,651) & (14,430) & \tabularnewline
Deductions (wife) & 7,623 & 7,565 & 0.01 &  & 8,108 & 8,193 & $-$0.01\tabularnewline
 & (6,344) & (6,358) &  &  & (9,454) & (9,546) & \tabularnewline
 &  &  &  &  &  &  & \tabularnewline
Labor income (wife) & 119,978 & 120,443 & $-$0.02 &  & 83,925 & 84,504 & $-$0.02\tabularnewline
 & (26,352) & (26,039) &  &  & (37,209) & (36,771) & \tabularnewline
\hline 
\end{tabular}
\medskip
\caption*{\footnotesize Notes: The table lists the means of covariates in 1986 by treatment status among workers employed in 1986 or 1993. Standard deviations are in parentheses. ``N. d." denotes a normalized difference, defined as $\frac{\overline{X}_{86} - \overline{X}_{93}}{\sqrt{(S_{86}^2 + S_{93}^2) / 2}}$ for each covariate, where $\overline{X}_{86}$ and $\overline{X}_{93}$ are the means among workers employed in 1986 and 1993 respectively, and $S_{86}$ and $S_{93}$ are the corresponding standard deviations. All monetary values are in Danish Krone (DKK), and DKK 1 in 1986 equals USD 0.3 in 2022. Children are defined as individuals younger than 17 years old. Low education is defined as completing primary education. Middle education is defined as completing high school or vocational education. High education is defined as holding a bachelor's, master's, or Ph.D. degree. Full-time jobs are defined as working more than 30 hours per week. The sample is males who, in 1986, were (i) younger than 50 years old and (ii) employed on the 28th of November. Furthermore, in 1986, (iii) they were married, and (iv) their wives had (strictly) positive labor income. The treated and control individuals are defined by the treatment assignment \eqref{eq:treated}. They are restricted to the low-income group ($120,000 \leq \text{LI}_{i86} < 160,000$).}
\end{table} 

\begin{table}[!htbp]\centering
\caption{Summary statistics of pre-reform covariates for the low-income and placebo groups}
\label{tab:CovariatePlacebo}
\begin{tabular}{lrrrcrrr}
\hline 
 & \multicolumn{3}{c}{Low-income} &  & \multicolumn{3}{c}{Placebo}\tabularnewline
\cline{2-4} \cline{6-8} 
Variable & Treated & Control & N. d. &  & Treated & Control & N. d.\tabularnewline
\hline 
\hline 
Age & 36.8 & 36.1 & 0.11 &  & 36.2 & 35.0 & 0.17\tabularnewline
 & (6.7) & (7.2) &  &  & (7.3) & (7.5) & \tabularnewline
Number of children & 1.3 & 1.5 & $-$0.11 &  & 1.5 & 1.5 & $-$0.03\tabularnewline
 & (0.9) & (0.9) &  &  & (0.9) & (1.0) & \tabularnewline
Low education (\%) & 35.4 & 38.6 & $-$0.07 &  & 38.5 & 41.6 & $-$0.06\tabularnewline
 & (47.8) & (48.7) &  &  & (48.7) & (49.3) & \tabularnewline
Middle education (\%) & 57.7 & 56.4 & 0.03 &  & 57.6 & 54.3 & 0.07\tabularnewline
 & (49.4) & (49.6) &  &  & (49.4) & (49.8) & \tabularnewline
High education (\%) & 6.9 & 5.0 & 0.08 &  & 3.9 & 4.2 & $-$0.01\tabularnewline
 & (25.4) & (21.8) &  &  & (19.3) & (20.0) & \tabularnewline
Full-time job (\%) & 46.7 & 46.3 & 0.01 &  & 45.2 & 39.2 & 0.12\tabularnewline
 & (49.9) & (49.9) &  &  & (49.8) & (48.8) & \tabularnewline
Private-sector job (\%) & 70.5 & 68.5 & 0.04 &  & 70.1 & 73.7 & $-$0.08\tabularnewline
 & (45.6) & (46.4) &  &  & (45.8) & (44.0) & \tabularnewline
 &  &  &  &  &  &  & \tabularnewline
Labor income & 148,596 & 142,386 & 0.65 &  & 142,815 & 141,994 & 0.08\tabularnewline
 & (8,202) & (10,625) &  &  & (10,612) & (10,620) & \tabularnewline
Capital income & $-$38,742 & $-$43,596 & 0.25 &  & $-$42,221 & $-$41,384 & $-$0.04\tabularnewline
 & (15,513) & (22,617) &  &  & (21,866) & (22,357) & \tabularnewline
Deductions & 10,647 & 10,354 & 0.04 &  & 9,948 & 10,655 & $-$0.10\tabularnewline
 & (6,605) & (7,903) &  &  & (7,263) & (7,621) & \tabularnewline
Capital income (wife) & $-$5,285 & $-$6,880 & 0.12 &  & $-$4,533 & $-$4,578 & 0.00\tabularnewline
 & (12,244) & (14,651) &  &  & (11,935) & (11,875) & \tabularnewline
Deductions (wife) & 7,623 & 8,108 & $-$0.06 &  & 6,267 & 4,909 & 0.28\tabularnewline
 & (6,344) & (9,454) &  &  & (5,640) & (3,849) & \tabularnewline
 &  &  &  &  &  &  & \tabularnewline
Labor income (wife) & 119,978 & 83,925 & 1.12 &  & 75,358 & 33,070 & 3.07\tabularnewline
 & (26,352) & (37,209) &  &  & (8,224) & (17,657) & \tabularnewline
\hline 
\hline 
Number of individuals & 11,642 & 28,338 &  &  & 7,086 & 7,083 & \tabularnewline
N. of observations (1981--1993) & 150,309 & 365,257 &  &  & 91,387 & 90,999 & \tabularnewline
\hline 
\end{tabular} 
\medskip
\caption*{\footnotesize Notes: The table lists the means of covariates in 1986 by treatment status for the low-income and placebo groups. Standard deviations are in parentheses. ``N. d." denotes a normalized difference, defined as $\frac{\overline{X}_T - \overline{X}_C}{\sqrt{(S_T^2 + S_C^2) / 2}}$ for each covariate, where $\overline{X}_T$ and $\overline{X}_C$ are the means among the treated and control individuals respectively, and $S_T$ and $S_C$ are the corresponding standard deviations. As a general guideline, a covariate is considered as balanced between the treated and control individuals if its normalized difference is less than 0.25. All monetary values are in Danish Krone (DKK), and DKK 1 in 1986 equals USD 0.3 in 2022. Children are defined as individuals younger than 17 years old. Low education is defined as completing primary education. Middle education is defined as completing high school or vocational education. High education is defined as holding a bachelor's, master's, or Ph.D. degree. Full-time jobs are defined as working more than 30 hours per week. The sample is males who, in 1986, were (i) younger than 50 years old and (ii) employed on the 28th of November. Furthermore, in 1986, (iii) they were married, and (iv) their wives had (strictly) positive labor income. The treated and control individuals are defined by the treatment assignment \eqref{eq:treated}. They are restricted to the low-income group ($120,000 \leq \text{LI}_{i86} < 160,000$). The placebo-treated and placebo-control individuals are defined by the placebo assignment \eqref{eq:placebo}.}
\end{table} 

\clearpage

\section*{Figures}

\begin{figure}[!htbp]\centering
\caption{Distributions of pre-reform labor income $\text{LI}_{i86}$}
\label{fig:LI_density}
\medskip
\includegraphics[scale=0.6]{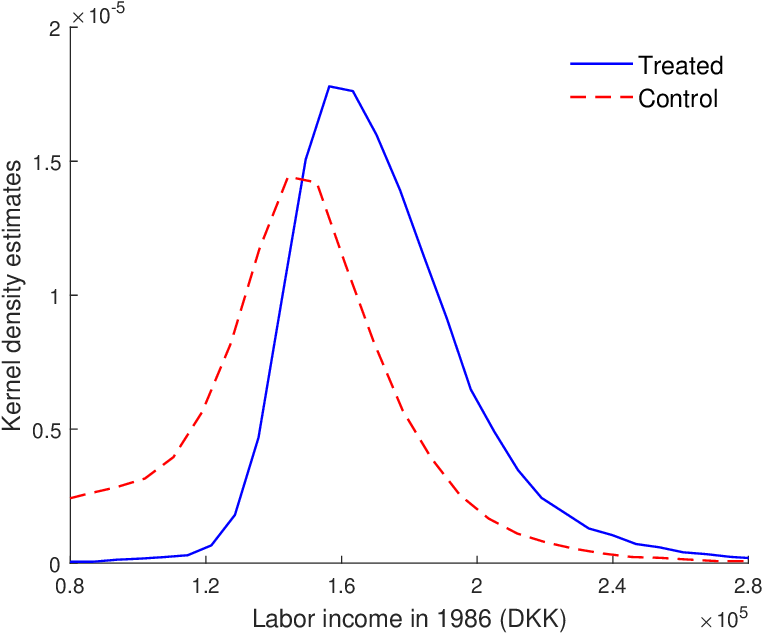}\hfill\includegraphics[scale=0.6]{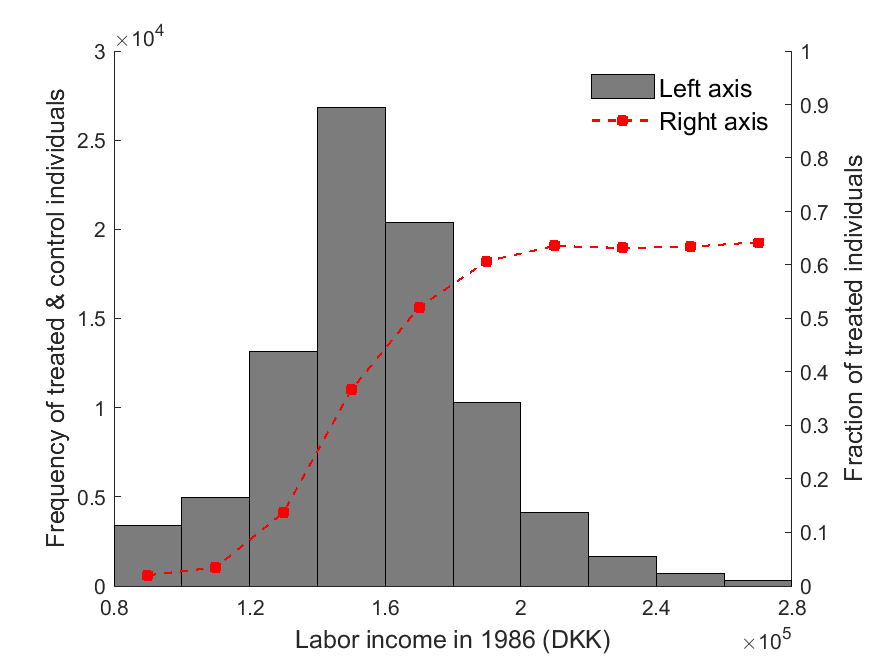} 
\medskip
\caption*{\footnotesize Notes: The figure presents the distributions of pre-reform labor income $\text{LI}_{i86}$. The left panel plots the kernel density estimates of $\text{LI}_{i86}$ by treatment status. The estimation is based on a \texttt{ksdensity} function in MATLAB with default settings. The right panel plots the histogram of $\text{LI}_{i86}$ with a bin width of DKK 20,000 among the treated and control individuals (on the left axis), and the fractions of the treated individuals within each bin (on the right axis). DKK 1 in 1986 equals USD 0.3 in 2022. The sample is males who, in 1986, were (i) younger than 50 years old and (ii) employed on the 28th of November. Furthermore, in 1986, (iii) they were married, and (iv) their wives had (strictly) positive labor income. The treated and control individuals are defined by the treatment assignment \eqref{eq:treated}.}
\end{figure}

\begin{figure}[!htbp]\centering
\caption{Bracket locations for the placebo group}
\label{fig:bracket_placebo}
\medskip
\includegraphics[scale=0.6]{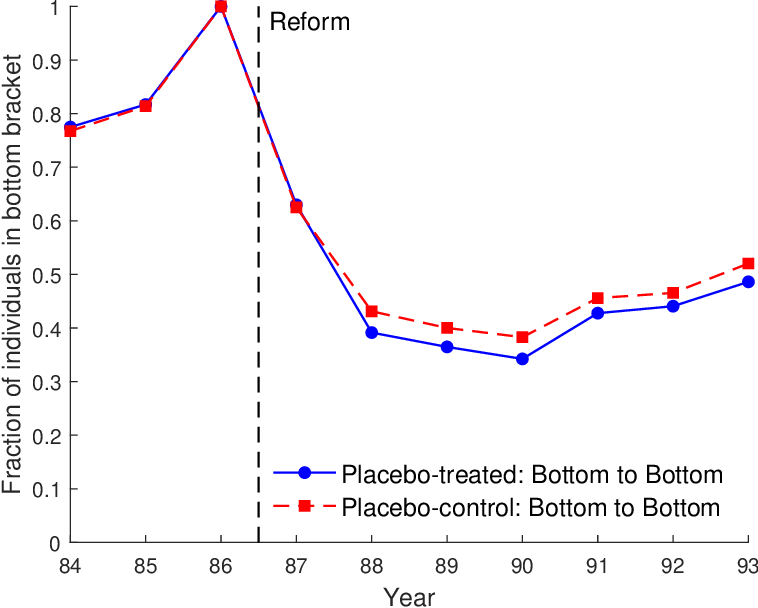}\hfill\includegraphics[scale=0.6]{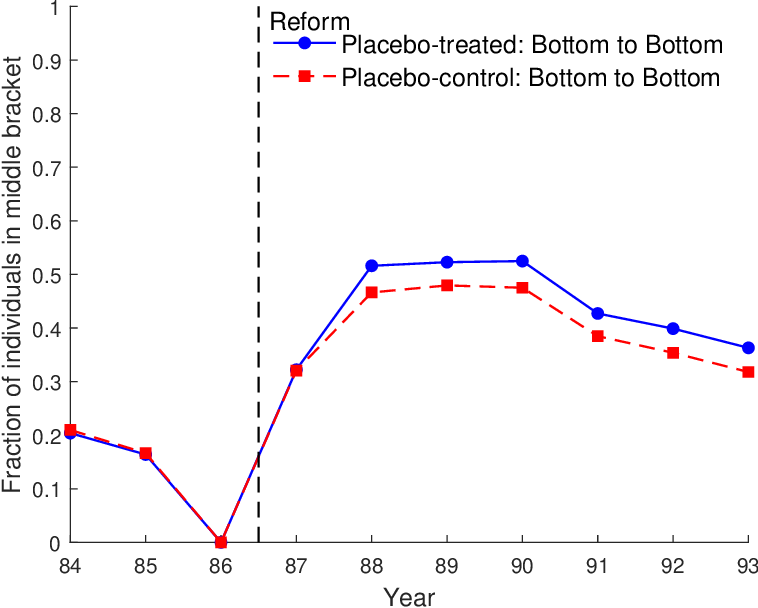}
\medskip
\caption*{\footnotesize Notes: The figure presents bracket locations for the placebo group. The left (right) panel plots the fractions of individuals located in the bottom (middle, respectively) bracket by treatment status. Bracket locations from 1981 to 1983 are missing due to data limitations. The sample is males who, in 1986, were (i) younger than 50 years old and (ii) employed on the 28th of November. Furthermore, in 1986, (iii) they were married, and (iv) their wives had (strictly) positive labor income. They are restricted to the low-income group ($120,000 \leq \text{LI}_{i86} < 160,000$). The placebo-treated and placebo-control individuals are defined by the placebo assignment \eqref{eq:placebo}.}
\end{figure}

\clearpage

\section*{Appendix}

\appendix
\counterwithin{figure}{section}
\counterwithin{table}{section}

\section{Background of the 1987 tax reform}\label{app:tax}
Like other Scandinavian countries, Denmark has a high tax burden. According to \citet{Kleven2014a}, its ratio of tax revenue to GDP in 2012 was 48 percent, higher than that in other developed countries such as Germany (36 percent), the United Kingdom (35 percent), and the United States (25 percent). Denmark collects approximately half of its revenue from individual income taxes.

Before the 1987 reform, the Danish income tax system was characterized by high marginal tax rates and narrow tax bases. As the left panel of Table \ref{tab:1987reform} shows, the top marginal tax rate reached 73 percent. Regarding the narrow tax bases, as Table \ref{tab:Covariate} shows, capital income was negative on average due to interest payments on debt, such as mortgage loans. This negative capital income thus narrowed the tax bases calculated as $\text{LI} + \text{CI} - \text{D}$.

Given these points, the reform was designed to broaden the tax bases and narrow the difference in marginal tax rates across the three brackets. First, it changed the tax bases for the middle and top brackets from $\text{LI} + \text{CI} - \text{D}$ to $\text{LI} + [\text{CI}>0]$ and $\text{LI} + [\text{CI}>60\text{k}]$, respectively. The reform thus broadened the tax bases by reducing the tax value of negative capital income and itemized deductions. Second, Figure \ref{fig:tax_func} in Appendix \ref{app:fig} plots marginal tax rates on labor income (LI) as a function of LI, before and after the 1987 reform. The tax rates and bracket cutoffs are taken from Table \ref{tab:1987reform}. For simplicity, we here assume single individuals with zero capital income and deductions ($\text{CI} = \text{D} = 0$). It is clear from the figure that the reform lowered the top and middle tax rates but raised the bottom tax rate, thereby narrowing the difference in marginal tax rates across the three brackets.\footnote{Although the reform flattened the tax schedule, it was approximately ex-ante revenue neutral by introducing green taxes levied on the consumption of natural resources.}

\section{Overview of the tax simulator}\label{app:simulator}
This appendix section explains the inputs of the simulator (including data sources), all income concepts necessary for the simulations, and the outputs of the simulator used in the empirical analysis.

\subsection{Inputs of the tax simulator}
The tax simulator takes as input information on the Danish income tax system (e.g., statutory tax rates) and information on individual income and demographic characteristics. The former information is primarily obtained from the website of the Danish Ministry of Taxation at \url{https://www.skm.dk}. The latter information is obtained from population-wide Danish administrative datasets. We refer to these datasets by their filenames on the server of Statistics Denmark used for the computations (ECONAU project 707275 via Aarhus University). The filenames are INDK, INDH, and PERSONER.

The three datasets are annual panels constructed from several registers (e.g., tax returns), cover all legal residents in Denmark aged 15--74 (on the 31st of December each year) since 1980, share a common individual ID, and contain extensive information. INDK and INDH contain administrative records on income tax assessments and public transfers, such as unemployment and sick leave benefits; we use variables regarding individual income and joint taxation. PERSONER contains information on demographic characteristics; we use variables regarding the municipality of residence, marital status, and the ID of his or her spouse (if married).

We construct a dataset for the simulations as follows. We first link INDK, INDH, and PERSONER using the common individual ID and a year variable; thus, the unit of observation is person-year. We then create variables regarding spousal income using the individual ID, his or her income, and the spousal ID. To this dataset, we next add information on regional taxes (e.g., statutory tax rates) using the municipal ID and a year variable. Information on national taxes is coded in the simulator.

\subsection{All income concepts}
The constructed dataset contains precise, individual-level measures of five income concepts necessary for the simulations. Table \ref{tab:IncomeDef} lists three key income concepts in the Danish income tax system: labor income (LI), capital income (CI), and itemized deductions (D). For accurate simulations, we need two additional income concepts of minor importance for our sample. The first is personal income, which includes labor income plus public transfers (e.g., unemployment and sick leave benefits) minus pension contributions. The second is stock income, which includes dividends and realized capital gains from shares \citep{Kleven2014b}. In the main text, we omit these two income concepts for ease of exposition.

\subsection{Outputs of the tax simulator}
Using all the necessary income concepts (LI, CI, ..., $\text{LI}^\text{w}$, $\text{CI}^\text{w}$, ...), demographic characteristics, and information on the Danish income tax system, we simulate bracket locations and effective marginal tax rates for individuals each year over the period 1984--1993. These outputs are missing from 1981 to 1983 due to data limitations. We describe and use the bracket locations in Section \ref{sec:emp_str} to define treated and control individuals. We describe and use the effective marginal tax rates in Section \ref{sec:DID} to compute elasticities. Finally, we link the dataset containing these input and output variables to IDA and job spell data using the common individual ID and a year variable (see Section \ref{sec:data} for details on IDA and job spell data).

\section{Distributions of pre-reform labor income among single males}\label{app:single}
Figure \ref{fig:LI_density_single} in Appendix \ref{app:fig} plots the kernel density estimates of pre-reform labor income $\text{LI}_{i86}$ by treatment status among single males. Except for marital status, the sample is the same as in Figure \ref{fig:LI_density}. The dashed line indicates a cutoff for the middle bracket under the inflation-adjusted 1987 tax system, which is consistent with $\widetilde{\text{M}}^{87}(z_{i86})$ in the treatment assignment \eqref{eq:treated}. Among single males, the two distributions do not sufficiently overlap due to a lack of variation to exploit; the treated individuals generally have higher $\text{LI}_{i86}$ and thus are mechanically pushed upward to the middle bracket under the 1987 tax system, i.e., $\widetilde{\text{M}}^{87}(z_{i86})$. Figures \ref{fig:LI_density} and \ref{fig:LI_density_single} demonstrate that by leveraging the joint taxation and variation in wives' income, we identify the treated and control individuals with the overlapping distributions of $\text{LI}_{i86}$.

Note that when two distributions do not sufficiently overlap, it is challenging to robustly control for pre-reform labor income $\text{LI}_{i86}$. In such cases, linear regression relies on extrapolation and becomes sensitive to the specifications of control variables \citep{Abadie2015, Imbens2015}. For example, the literature on the elasticity of taxable income often finds estimates sensitive to the specifications of pre-reform income. This sensitivity occurs because researchers often compare a certain income group affected by a tax reform to an unaffected, higher- or lower-income group \citep{Saez2012}.

\section{Computation of outcome variables: wages and hours}\label{app:variable}
\paragraph{Gross hourly wages.}
Gross hourly wages in IDA are computed as annual earnings from a November job divided by annual hours worked for that job. Note that labor income (LI) includes annual earnings from both November and non-November jobs.

The annual earnings (i.e., the numerator) are reported to the tax authorities by employers for income tax purposes and are subject to minimal misreporting and measurement errors. These annual earnings include regular pay, overtime pay, bonuses, vacation pay, and illness allowances, but not employer pension contributions.

The annual hours (i.e., the denominator) in IDA are estimated from annual pension contribution records (known as ATP for ``Arbejdsmarkedets Till{\ae}gspension") by leveraging the fact that mandatory employer contributions to a supplementary pension scheme depend only on hours worked by individual employees \citep{Lund2016}. These annual hours do not include overtime work, vacation, and periods of absence due to illness; among these missing components, overtime work will be the most important.

We argue that the lack of overtime work is not a serious concern for three reasons. First, \citet{Lund2016} document that the estimated hourly wages are precise by comparing them to hourly wages obtained from another register called ``L{\o}nstatistik" (Wage and Salary Statistics). Second, \citet{Hummels2014} utilize data on overtime work obtained from IDA in 2006 and document that annual hours including overtime work are highly correlated with those excluding overtime work. Third, the lack of overtime work will pose a threat to identification in the DID design if overtime work correlates with both outcome dynamics (e.g., wage dynamics) and our instrumental variable (i.e., wives' labor income $\text{LI}_{i86}^\text{w}$)---a violation of the exclusion restriction. However, in Section \ref{sec:DID}, we provide evidence against such a correlation by showing parallel pre-reform outcome dynamics and by conducting a placebo test.

\paragraph{Daily hours worked.}
Daily hours worked in the ``November-job" data are computed as annual hours worked for a November job divided by annual days worked for that job. The annual hours (i.e., the numerator) are estimated similarly to the annual hours in IDA (described above); indeed, they are highly correlated. The annual days (i.e., the denominator) are determined from the start and end dates of each job.\footnote{We cannot use IDA to compute daily hours worked because it does not contain the start and end dates of jobs.}

\section{Additional figures}\label{app:fig}

\begin{figure}[!htbp]\centering
\caption{Overview of the 1987 tax reform}
\label{fig:tax_func}
\medskip
\includegraphics[scale=0.8]{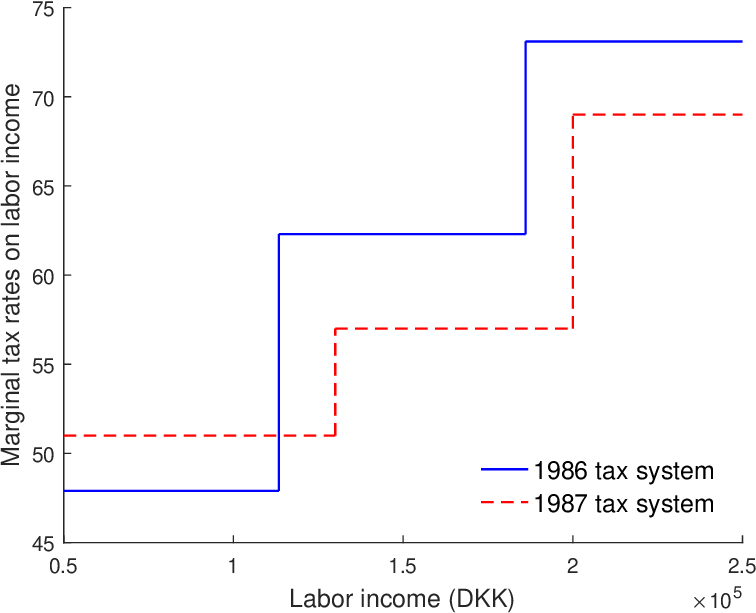}
\medskip
\caption*{\footnotesize Notes: The figure plots marginal tax rates on labor income (LI) as a function of LI, before and after the 1987 reform. The tax rates and bracket cutoffs are taken from Table \ref{tab:1987reform}. For simplicity, we here assume single individuals with zero capital income and deductions ($\text{CI} = \text{D} = 0$). DKK 1 in 1986 equals USD 0.3 in 2022.}
\end{figure}

\begin{figure}[!htbp]\centering
\caption{Distributions of pre-reform labor income $\text{LI}_{i86}$ (single males)}
\label{fig:LI_density_single}
\medskip
\includegraphics[scale=0.8]{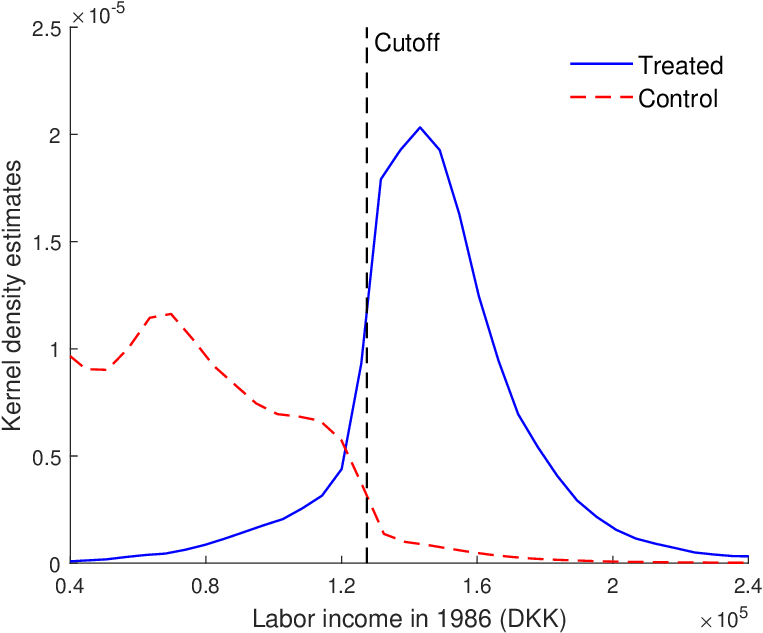}
\medskip
\caption*{\footnotesize Notes: The figure plots the kernel density estimates of pre-reform labor income $\text{LI}_{i86}$ by treatment status among single males. Except for marital status, the sample is the same as in Figure \ref{fig:LI_density}. The dashed line indicates a cutoff for the middle bracket under the inflation-adjusted 1987 tax system. The estimation is based on a \texttt{ksdensity} function in MATLAB with default settings. DKK 1 in 1986 equals USD 0.3 in 2022. The sample is males who, in 1986, were (i) younger than 50 years old and (ii) employed on the 28th of November. The treated and control individuals are defined by the treatment assignment \eqref{eq:treated}.}
\end{figure}

\begin{figure}[!htbp]\centering
\caption{Bracket locations for the low-income group (other brackets)}
\label{fig:bracket_low_app}
\medskip
\includegraphics[scale=0.6]{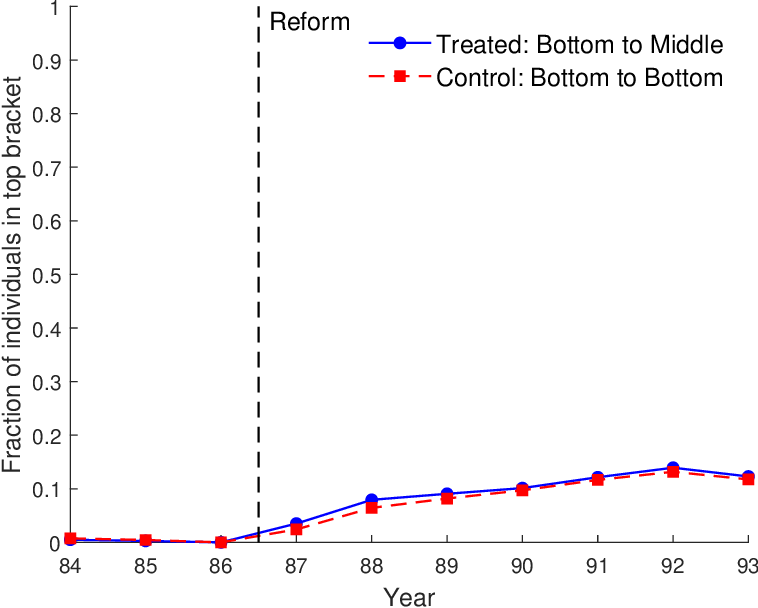}\hfill\includegraphics[scale=0.6]{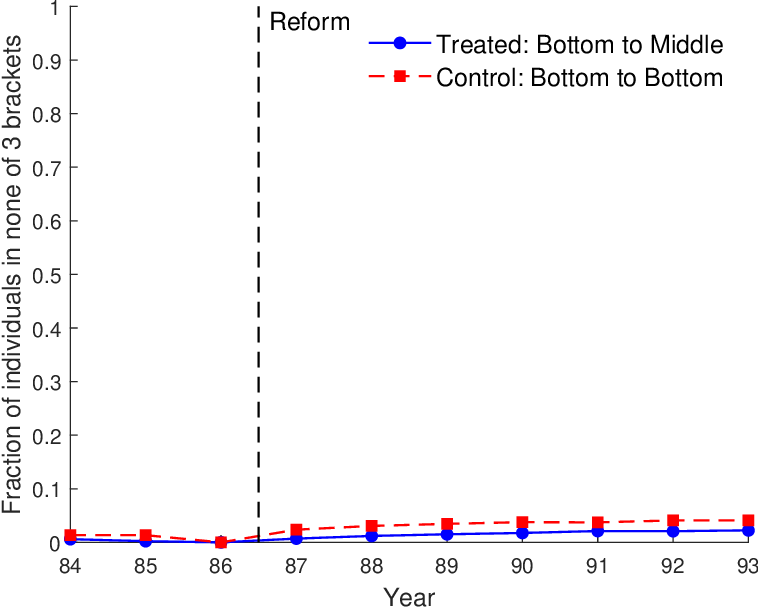}
\medskip
\caption*{\footnotesize Notes: The figure presents bracket locations for the low-income group ($120,000 \leq \text{LI}_{i86} < 160,000$). The left (right) panel plots the fractions of individuals located in the top bracket (none of the three brackets, respectively) by treatment status. Bracket locations from 1981 to 1983 are missing due to data limitations. The sample is males who, in 1986, were (i) younger than 50 years old and (ii) employed on the 28th of November. Furthermore, in 1986, (iii) they were married, and (iv) their wives had (strictly) positive labor income. The treated and control individuals are defined by the treatment assignment \eqref{eq:treated}.}
\end{figure}

\begin{figure}[!htbp]\centering
\caption{Bracket locations for the placebo group (other brackets)}
\label{fig:bracket_placebo_app}
\medskip
\includegraphics[scale=0.6]{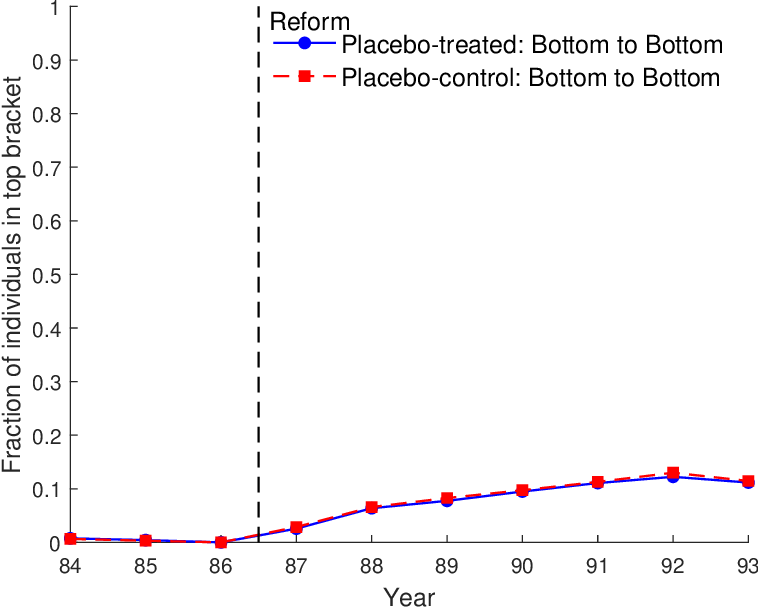}\hfill\includegraphics[scale=0.6]{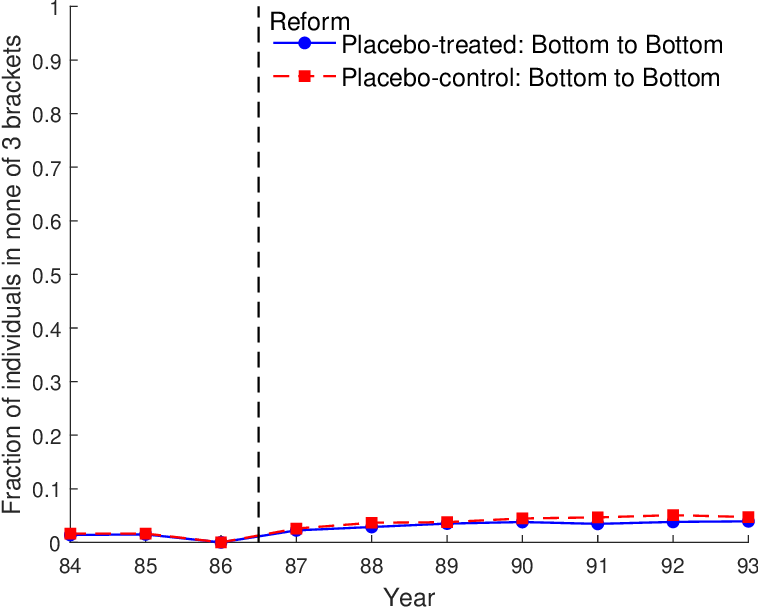}
\medskip
\caption*{\footnotesize Notes: The figure presents bracket locations for the placebo group. The left (right) panel plots the fractions of individuals located in the top bracket (none of the three brackets, respectively) by treatment status. Bracket locations from 1981 to 1983 are missing due to data limitations. The sample is males who, in 1986, were (i) younger than 50 years old and (ii) employed on the 28th of November. Furthermore, in 1986, (iii) they were married, and (iv) their wives had (strictly) positive labor income. They are restricted to the low-income group ($120,000 \leq \text{LI}_{i86} < 160,000$). The placebo-treated and placebo-control individuals are defined by the placebo assignment \eqref{eq:placebo}.}
\end{figure}

\end{document}